\documentclass[twocolumn,times,modern]{aastex62}
\pdfoutput=1

\usepackage{xspace}
\usepackage{mathptmx}
\usepackage{natbib}
\usepackage{booktabs}
\usepackage{amsmath}

\newcommand{\fif}{4U~1538$-$522\xspace}
\newcommand{\msol}{$M_{\sun}$\xspace}

\newcommand{\qvnor}{QV~Nor\xspace}
\newcommand{\suz}{\emph{Suzaku}\xspace}
\newcommand{\nustar}{\emph{NuSTAR}\xspace}
\newcommand{\integral}{\emph{INTEGRAL}\xspace}
\newcommand{\chandra}{\emph{Chandra}\xspace}
\newcommand{\hetgs}{HETGS\xspace}
\newcommand{\rxte}{\emph{RXTE}\xspace}
\newcommand{\nh}{$N_{\mathrm{H}}$\xspace}
\newcommand{\nhpc}{$N_{\rm H,pc}$\xspace}
\newcommand{\fpc}{$f_{\rm pc}$\xspace}

\newcommand{\heaver}{v6.22\xspace}
\newcommand{\caldbver}{20170727\xspace}
\newcommand{\isisver}{v1.6.2-40\xspace}
\newcommand{\pflx}{ph\,cm$^{-2}$\,s$^{-1}$\xspace}
\newcommand{\eflx}{erg\,cm$^{-2}$\,s$^{-1}$\xspace}

\begin{document}

\title{The First \nustar Observation of \fif: Updated Orbital Ephemeris and A Strengthened Case for an Evolving Cyclotron Line Energy}

\author[0000-0002-1676-6954]{Paul~B.~Hemphill}
\affiliation{Kavli Institute for Astrophysics and Space Research, Massachusetts Institute of Technology, Cambridge, MA 02139, USA}
\email{pbh@space.mit.edu}
\author{Richard~E.~Rothschild}
\affiliation{Center for Astrophysics and Space Sciences, University of California, San Diego, 9500 Gilman Dr., La Jolla, CA 920093-0424, USA}
\author{Diana~M.~Cheatham}
\affiliation{Center for Space Science and Technology, University of Maryland Baltimore County, 1000 Hilltop Circle, Baltimore, MD 21250, USA}
\affiliation{CRESST and NASA Goddard Space Flight Center, Astrophysics Science Division, Code 661, Greenbelt, MD 20771, USA}
\author{Felix~F\"{u}rst}
\affiliation{European Space Astronomy Centre (ESAC), Science Operations Departement, 28692 Villanueva de la Cañada, Madrid, Spain}
\author{Peter~Kretschmar}
\affiliation{European Space Astronomy Centre (ESAC), Science Operations Departement, 28692 Villanueva de la Cañada, Madrid, Spain}
\author{Matthias~K\"{u}hnel}
\affiliation{Dr.\ Karl Remeis-Sternwarte \& Erlangen Center for Astroparticle Physics, Sternwartstr. 7, 96049 Bamberg, Germany}
\author{Katja~Pottschmidt}
\affiliation{Center for Space Science and Technology, University of Maryland Baltimore County, 1000 Hilltop Circle, Baltimore, MD 21250, USA}
\affiliation{CRESST and NASA Goddard Space Flight Center, Astrophysics Science Division, Code 661, Greenbelt, MD 20771, USA}
\author{R\"{u}diger~Staubert}
\affiliation{Institut f\"{u}r Astronomie und Astrophysik, Universit\"{a}t T\"{u}bingen, Sand 1, 72076 T\"{u}bingen, Germany}
\author{J\"{o}rn~Wilms}
\affiliation{Dr.\ Karl Remeis-Sternwarte \& Erlangen Center for Astroparticle Physics, Sternwartstr. 7, 96049 Bamberg, Germany}
\author{Michael~T.~Wolff}
\affiliation{Space Science Division, Naval Research Laboratory, Washington, DC 20375-5352, US}

\begin{abstract}
  We have performed a comprehensive spectral and timing analysis of the first
  \nustar observation of the high-mass X-ray binary \fif. The observation
  covers the X-ray eclipse of the source, plus the eclipse ingress and egress.
  We use the new measurement of the mid-eclipse time to update the orbital
  parameters of the system and find marginally-significant evolution in the orbital
  period, with $\dot{P}_{\rm orb}/P_{\rm orb} = \left(-0.95 \pm 0.37\right)
  \times 10^{-6}$\,yr$^{-1}$. The cyclotron line energy is found approximately
  1.2\,keV higher than \rxte measurements from 1997--2003, in line with the
  increased energy observed by \suz in 2012 and strengthening the case for
  secular evolution of \fif's CRSF. We additionally characterize the behavior
  of the iron fluorescence and emission lines and line-of-sight absorption as
  the source moves into and out of eclipse.
\end{abstract}


\section{Introduction}\label{sec:intro}

High-mass X-ray binaries (HMXBs) present interesting laboratories for a number
of astrophysical questions. As endpoints of stellar evolution, they provide
lever arms on models for stellar and binary evolution, and their high accretion
rates and strong magnetic fields (in HMXBs hosting neutron stars) allow us to
probe areas of plasma physics inaccessible to terrestrial laboratories. The
eclipsing X-ray pulsar \fif has been an interesting source in both of these
regards since its discovery by \textit{Uhuru} in the 1970s
\citep{giacconi_third_1974}. \textit{Uhuru} established the eclipsing nature of
the source \citep[see, e.g.,][]{cominsky_multimission_1991}, while X-ray pulsations were
discovered by \citet{davison_binary_1977} and \citet{becker_a1540-53_1977}. The
pulsar is slow-spinning, with a pulse period of $\sim$526.5\,s, and is
currently spinning down at a rate of $\sim$0.1\,s\,yr$^{-1}$ as revealed by
\textit{Fermi}/GBM\footnote{See
\url{http://gammaray.nsstc.nasa.gov/gbm/science/pulsars}}
\citep{finger_long-term_2009}. The distance is somewhere in the neighborhood of
6.5\,kpc \citep[see,
e.g.,][]{crampton_optical_1978,ilovaisky_optical_1979,reynolds_optical_1992,clark_chandra_2004};
\textit{Gaia} parallax measurements \citep{bailer-jones_estimating_2018} place the source at
$6.6^{+2.2}_{-1.5}$\,kpc. In this work we adopt the 6.4\,kpc distance measured
by \citet{reynolds_optical_1992}, as we have in previous work on this source.
The X-ray pulsar accretes from the stellar wind of the B0Iab supergiant \qvnor
\citep{reynolds_optical_1992}, which it orbits every 3.73\,d with regular
eclipses
\citep{clark_orbit_2000,mukherjee_orbital_2006,baykal_recent_2006,falanga_ephemeris_2015}.
The neutron star appears to be very low-mass, at $\sim$1\,\msol
\citep{rawls_refined_2011,falanga_ephemeris_2015}. Despite the eclipsing
nature of the source, the binary orbit has historically been difficult to nail
down. The orbit is probably eccentric, at $e \approx 0.18$
\citep{clark_orbit_2000,mukherjee_orbital_2006}, although
\citet{clark_orbit_2000} and \citet{baykal_recent_2006} also address the
circular case, and \citet{rawls_refined_2011} reported difficulty reconciling
the observed eclipse duration with an eccentric orbit. The source also may
exhibit apsidal advance, as reported by \citet{falanga_ephemeris_2015},
although at $1.3^{\circ} \pm 0.6$\,yr$^{-1}$ this is poorly constrained.

In the area of strong magnetic fields and accretion physics, \fif is notable
for having a cyclotron resonance scattering feature (CRSF, or ``cyclotron
line'') in its X-ray spectrum, at $\sim$22\,keV. These pseudo-absorption
features arise due to the quantization of cyclotron motion for electrons in
strong magnetic fields, which leads to a scattering cross section for photons
that is strongly dependent on energy, direction, and polarization. These features
provide the only direct measurement of the magnetic field of a neutron star
(although it should be noted that this measures a \textit{local} field strength,
compared to more global measurements from, e.g., radio pulsar spin-down).
Approximately three dozen sources now have established cyclotron lines,
starting with the first discovered in Her~X-1 by \citet{trumper_evidence_1978}.
The recent review by \citet{staubert_cyclotron_2018} provides an excellent
overview of the current state of the art in cyclotron line studies.

A great deal of work in the past decade has been concerned with the variability
of cyclotron lines, mostly in terms of the line energy's relation to the X-ray
luminosity of the source. The basic picture until a few years ago was that
sources at very high luminosities exhibited an anti-correlation between CRSF
energy and luminosity \citep[see the case of V~0332+53
by][]{tsygankov_completing_2010}, while lower-luminosity sources displayed a
positive correlation or no correlation \citep[see,
e.g.,][]{staubert_discovery_2007,yamamoto_discovery_2011,klochkov_outburst_2012,muller_0535_2013,hemphill_measurements_2013,furst_nustar_2014,hemphill_evidence_2016,doroshenko_luminosity_2017}.
Many sources also display considerable variability of the CRSF with pulse phase
\citep[see, e.g.,][]{hemphill_clumpy_2014}. Some sources display more complex
behavior, such as different variability depending on timescale \citep[e.g.,
A~0535+26; see][]{caballero_preoutburst_2008},  or possibly dependence on
super-orbital variability \citep[e.g., Her~X-1;
see][]{staubert_long-term_2014}. Neutron stars at high accretion rates present
a set of incredibly difficult theoretical problems, and their response to changes in
accretion rate, as well as the basic question of where cyclotron lines form in
the first place, has been a matter of some debate \citep[see,
e.g.,][]{becker_spectral_2012,poutanen_reflection_2013,mushtukov_critical_2015,mushtukov_positive_2015,nishimura_properties_2015,schwarm_cyclotron_2017,schwarm_cyclotron_2017a}.

This already-difficult problem has been complicated in recent years by the
observation of significant secular changes in cyclotron line energies
\textit{independent} of luminosity. The first-discovered instance of a secular
change in CRSF energy was Her~X-1, when
\citet{staubert_long-term_2014} reported the discovery of a long-term decay in
the line energy over the previous two decades. This trend continued for the next few
years \citep{staubert_continued_2016}, but recently,
\citet{staubert_inversion_2017} have presented evidence that the trend has
reversed.  \citet{laparola_swiftbat_2016} have argued that a similar decay in
CRSF energy is present in Vela~X-1, based on \textit{Swift}/BAT observations.
Finally, the BeXRB V~0332+53 displayed hysteresis in its CRSF evolution over
the course of its 2015 giant outburst (despite there being no sign of this
behavior during its 2003--2004 outburst), with a markedly lower CRSF energy at
the end of the outburst compared to the beginning
\citep{cusumano_unexpected_2016}; furthermore, the CRSF energy seems to have
rebounded back to its pre-outburst state by the time of its 2016 outburst
\citep{doroshenko_luminosity_2017,vybornov_changes_2018}.

\fif is the fourth source in this regard. Its CRSF was originally discovered by
\citet{clark_discovery_1990} in \textit{Ginga} observations, analyzed further by
\citet{mihara_observational_1995}, at $\sim$20\,keV. In
\citet{hemphill_evidence_2016}, we carried out an extensive re-analysis of the
available \rxte, \integral, and \suz data, concluding that there was evidence
that the CRSF energy had increased by $\sim$1.5\,keV between the \rxte
observations of the early 2000s and the 2012 \suz observation. A recent study
of \textit{AstroSat} data by \citet{varun_probing_2018} has also observed an
increased CRSF energy. This rise in energy runs opposite to the long-term
secular decays observed in Her~X-1 and Vela~X-1. It also differs from the
outburst-to-outburst increase in CRSF energy observed in V~0332+53 --- in \fif,
the change in CRSF energy does not appear to be associated with any major
changes in accretion rate, and operates on a timescale of decades, compared to
the $\sim$1\,yr timescale in V~0332+53.  \fif's CRSF also displays no detected
correlation with luminosity \citep{hemphill_evidence_2016}, unlike the other
sources with detected secular trends.

In this paper we examine our 2016 \nustar observation of \fif in detail.
Section \ref{sec:obs} summarizes the observation and data reduction. In
Section~\ref{sec:timing} we present an updated orbital ephemeris based on the
new mid-eclipse time. Sections \ref{sec:pre_results} and \ref{sec:ecl}
summarize the analyses of the pre-eclipse and eclipse portions of the
observation. Section~\ref{sec:discussion} discusses these results in the
context of past studies of this source and neutron star HMXBs in general.

\section{Observations and data reduction} \label{sec:obs}

\nustar \citep{harrison_nuclear_2013} observed \fif on 11 August,
2016 for a total exposure of 43.8\,ks. The full length of the observation is
84.7\,ks, and it covers the eclipse ingress, the eclipse, and the eclipse
egress. Based on our ephemeris and mid-eclipse time found in
Section~\ref{sec:timing}, and placing phase 0 at the mid-eclipse, the
observation covers orbital phases 0.84--0.1, with the eclipse extending
0.07--0.08\ in orbital phase to either side of the eclipse center.

\begin{figure*}
  \centering
  \plotone{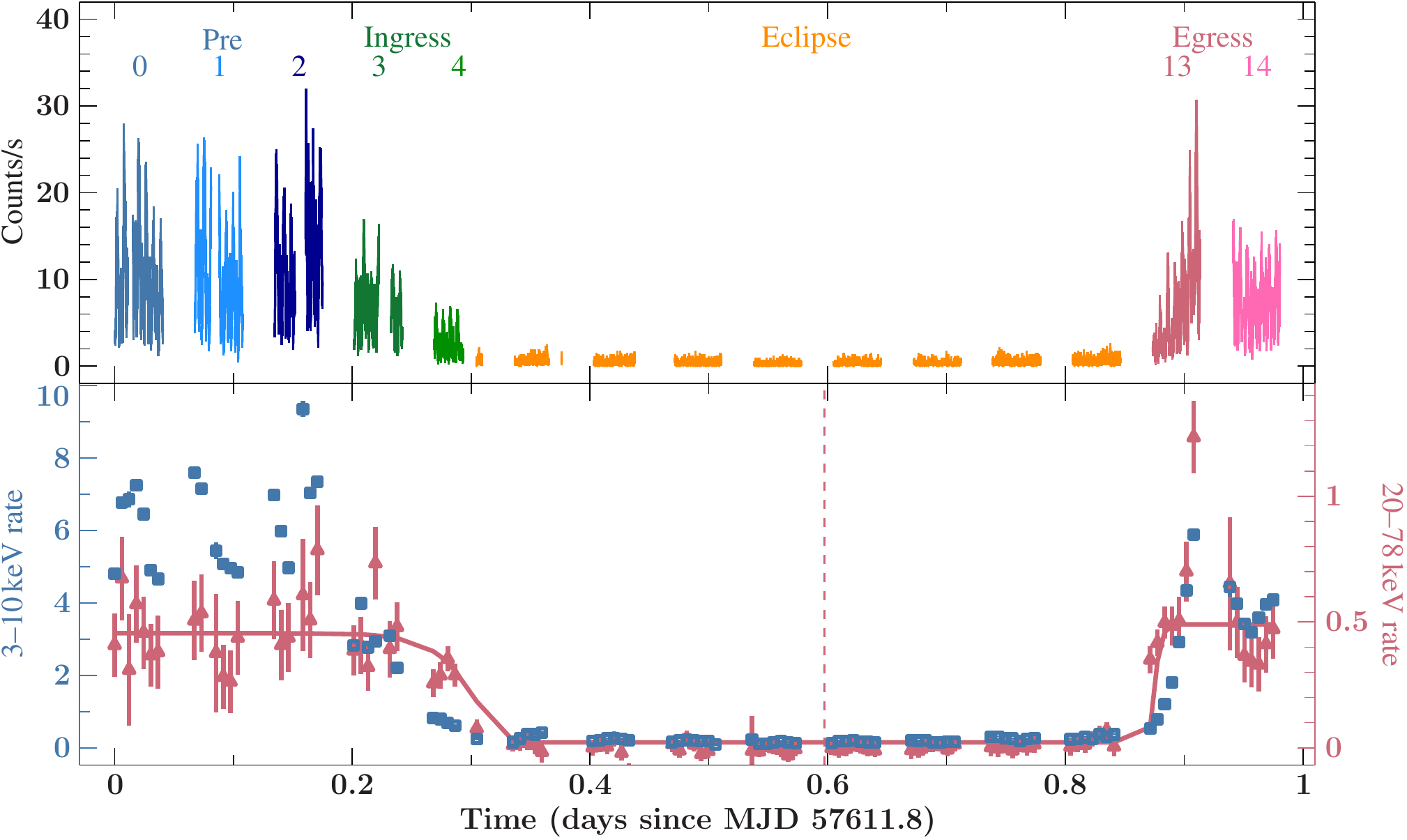}
  \caption{Upper plot displays the 3--78\,keV \nustar lightcurve of \fif. The
    pre-eclipse phase is colored in blue, the eclipse ingress in green, the
    eclipse in yellow, and the eclipse egress in red. Individual spectra for
    time-resolved spectroscopy are numbered (note that spectra 5--12 are the
    eclipse and are not investigated individually). Lower plot displays the
    background-subtracted 3--10 (blue squares) and 20--78\,keV (red
    triangles) lightcurves, binned at the pulse profile. The eclipse profile
    (see Equation~\ref{eqn:eclipse} and accompanying text) is overplotted in
    red, and the mid-eclipse time is marked with a dashed line.}
  \label{fig:lc}
\end{figure*}

\begin{figure*}
  \centering
  \plotone{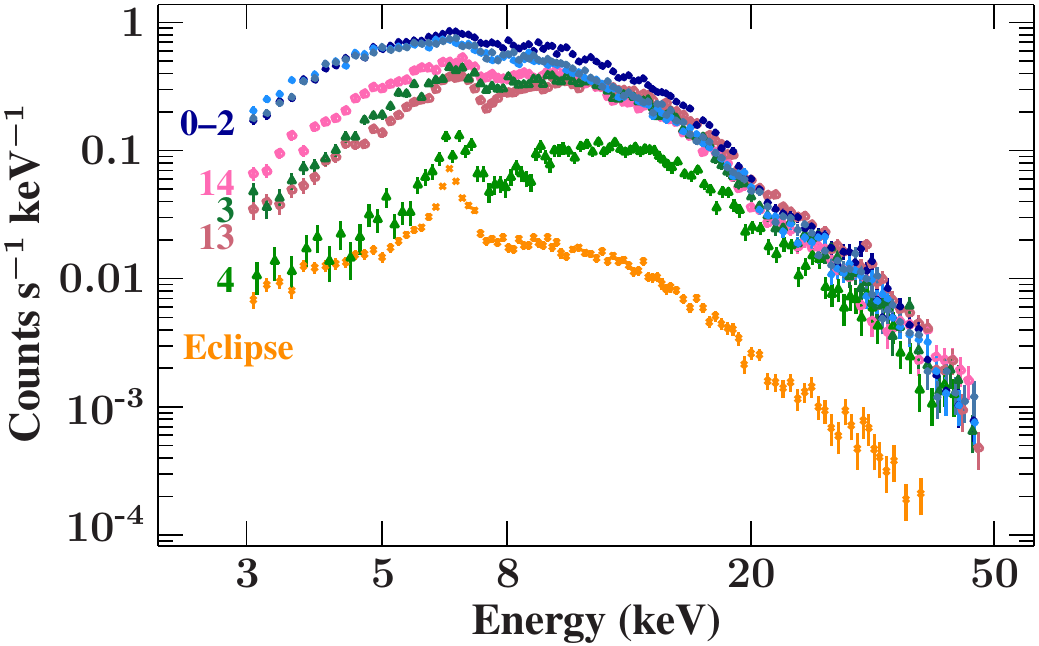}
  \caption{The \nustar spectra for the time intervals indicated in
    Figure~\ref{fig:lc}. FPMA and FPMB have been combined and rebinned
    for clarity. Spectra 0--2 (blue points), are the spectra selected for the
    pre-eclipse analysis. The eclipse ingress and egress are spectra 3--4
    (green triangles) and 13--14 (pink circles), respectively, while the
    eclipse spectrum is plotted in yellow crosses. The change in absorption
    over the observation is apparent.}
  \label{fig:spectra}
\end{figure*}

We extracted barycentered lightcurves and spectra using the standard \nustar
pipeline in HEASOFT \heaver and \nustar CALDB version \caldbver, using
circular 60\arcsec source and background regions. During our spectral analysis,
we verified that using a different background region did not have a significant
effect on our results. The lightcurve of the observation is plotted
in the upper panel of Figure~\ref{fig:lc}. The X-ray eclipse and the regular
Earth occultations in the observation provide the breakpoints for our initial
extraction of spectra, which are plotted in Figure~\ref{fig:spectra}. Based on
the qualitative appearance of these spectra, we define spectra 0--2 as the
``pre-eclipse'' spectra, 3--4 as the eclipse ingress, and 13--14 as the eclipse
egress, with the remainder making up the eclipse. We accordingly extracted the
average spectra for these intervals.

All analysis was carried out using the Interactive Spectral Interpretation
System \citep[ISIS;][]{houck_isis:_2000}, \isisver, and all uncertainties are
quoted at the 90\% level for a single parameter of interest unless otherwise
specified.

\section{Lightcurve analysis: updated orbital ephemeris} \label{sec:timing}

The background-subtracted 3--10 and 20--78\,keV lightcurves are plotted
in the lower panel of Figure~\ref{fig:lc}, where we have binned the lightcurves
at the pulse period (see Section~\ref{sec:pulse}) to filter out the variability
due to the source's pulsations. The harder band, largely unaffected by
absorption, shows a much sharper eclipse profile than the softer band. The
profile in the hard band is still moderately asymmetric, with a more gentle
ingress and very sharp egress, but it is not clear whether this is due to
source variability or the particular arrangement of Earth occultation and SAA
passage gaps in this observation.

\subsection{Eclipse duration} \label{sec:ecl_dur}

In order to find the eclipse duration and the mid-eclipse time, we use the
background-subtracted 20--78\,keV lightcurve rebinned at the pulse
period. Taking the orbital ephemeris from \citet{falanga_ephemeris_2015} and an
initial by-eye estimate of the mid-eclipse time, we computed the orbital phase
for each time bin in the lightcurve relative to the eclipse center, and applied
a model of the form $F_{\rm in}(\varphi) + F_{\rm out}(\varphi) + f_{\rm ecl}$
to the lightcurve, where $f_{\rm ecl}$ is the countrate during eclipse and
$F_{\rm in}(\varphi)$ and $F_{\rm out}(\varphi)$ are both the
double-exponential model employed by \citet{falanga_ephemeris_2015} to find the
eclipse ingress and egress phases:
\begin{equation} \label{eqn:eclipse}
  F\left(\varphi\right) = F_{0} \exp\left[\ln\left(0.01\right)\exp\left(\pm\frac{\varphi - \varphi{\rm e}}{\tau_{\rm e}}\right)\right],
\end{equation}
where the argument of the inner exponential is positive for the eclipse ingress
and negative for the eclipse egress. Here, $F_{0}$ is the pre- or
post-eclipse average countrate, $\varphi{\rm e}$ is the phase of eclipse ingress
or egress, and $\tau_{\rm e}$ is the transition phase width of the ingress or
egress. The results of this fit are presented in Table~\ref{tab:ecl}.

% WITHOUT BACKGROUND SUBTRACTION
% \begin{deluxetable}{lrr}
%   \tabletypesize{\footnotesize}
%   \tablewidth{0pt}
%   \tablecaption{Orbital phases and times for the eclipse ingress and egress,
%     eclipse duration, semi-eclipse angle, and mid-eclipse time, based on fits
%     to the eclipse profile with Equation~\ref{eqn:eclipse}. Uncertainties are
%     1$\sigma$ and upper limits are 99\%.\label{tab:ecl}}
%   \tablehead{\colhead{} & \colhead{Phase} & \colhead{Days}}
%   \startdata
%   $\varphi_{\rm in}$ & $0.930 \pm 0.002$ & $57612.146 \pm 0.007$\,MJD \\
%   $\tau_{\rm in}$ & $\left(5.8^{+1.2}_{-1.0}\right)\times 10^{-3}$ & $0.022 \pm 0.004$\,d \\
%   $\varphi_{\rm out}$ & $0.073 \pm 0.002$ & $57612.676 \pm 0.007$\,MJD \\
%   $\tau_{\rm out}$ & $< 3.8 \times 10^{-3}$ & $< 0.014$\,d \\
%   Eclipse duration & $0.142 \pm 0.003$ & $0.529 \pm 0.011$\,d \\
%   \midrule
% $\theta_{\rm e}$ & & $25.6^{\circ} \pm 0.5$ \\
%   $T_{\rm ecl}$ & & $57612.411 \pm 0.011$\,MJD
%   \enddata
% \end{deluxetable}

% WITH BACKGROUND SUBTRACTION
\begin{deluxetable}{lrr}
  \tabletypesize{\footnotesize}
  \tablewidth{0pt}
  \tablecaption{Orbital phases and times for the eclipse ingress and egress,
    eclipse duration, semi-eclipse angle, and mid-eclipse time, based on fits
    to the eclipse profile with Equation~\ref{eqn:eclipse}. Uncertainties are
    1$\sigma$ and upper limits are 99\%.\label{tab:ecl}}
  \tablehead{\colhead{} & \colhead{Phase} & \colhead{Days}}
  \startdata
  $\varphi_{\rm in}$ & $0.928 \pm 0.002$ & $57612.131 \pm 0.007$\,MJD \\
  $\tau_{\rm in}$ & $\left(5.3^{+1.5}_{-0.9}\right)\times 10^{-3}$ & $0.020^{+0.006}_{-0.003}$\,d \\
  $\varphi_{\rm out}$ & $0.071 \pm 0.003$ & $57612.665 \pm 0.010$\,MJD \\
  $\tau_{\rm out}$ & $< 3.8 \times 10^{-3}$ & $< 0.014$\,d \\
  Eclipse duration & $0.143 \pm 0.004$ & $0.533 \pm 0.015$\,d \\
  \midrule
  $\theta_{\rm e}$ & & $25.7^{\circ} \pm 1.1$ \\
  $T_{\rm ecl}$ & & $57612.401 \pm 0.013$\,MJD
  \enddata
\end{deluxetable}

Our eclipse duration of $0.529 \pm 0.011$\,d and semi-eclipse angle of
$25.6^{\circ} \pm 0.5$ are somewhat higher than the \integral eclipses reported
by \citet{falanga_ephemeris_2015}. The transition widths we measure for
eclipse ingress and egress are smaller than those of
\citet{falanga_ephemeris_2015}. These differences may simply be due to our
relatively sparse sampling compared to \citeauthor{falanga_ephemeris_2015}, as
we need to average over pulsations and lack data during Earth occults, or due
to eclipse-to-eclipse variability, which is averaged out in
\citeauthor{falanga_ephemeris_2015}'s folded eclipse profiles. Our values are
quite consistent with previous measurements of the eclipse duration found using
single observations
\citep[e.g.,][]{becker_a1540-53_1977,davison_regular_1977,makishima_spectra_1987}.
The eclipse width measured by eye, which is more comparable to the method used
in these previous works, is $0.560 \pm 0.014$\,d (estimating the uncertainty
from the widths of the gaps on either side of the eclipse), which is also
consistent with the older studies and longer than the eclipse measured by
\citet{falanga_ephemeris_2015}.

\subsection{A new orbital ephemeris} \label{sec:orbit}

\begin{figure}
  \centering
  \includegraphics[width=0.5\textwidth]{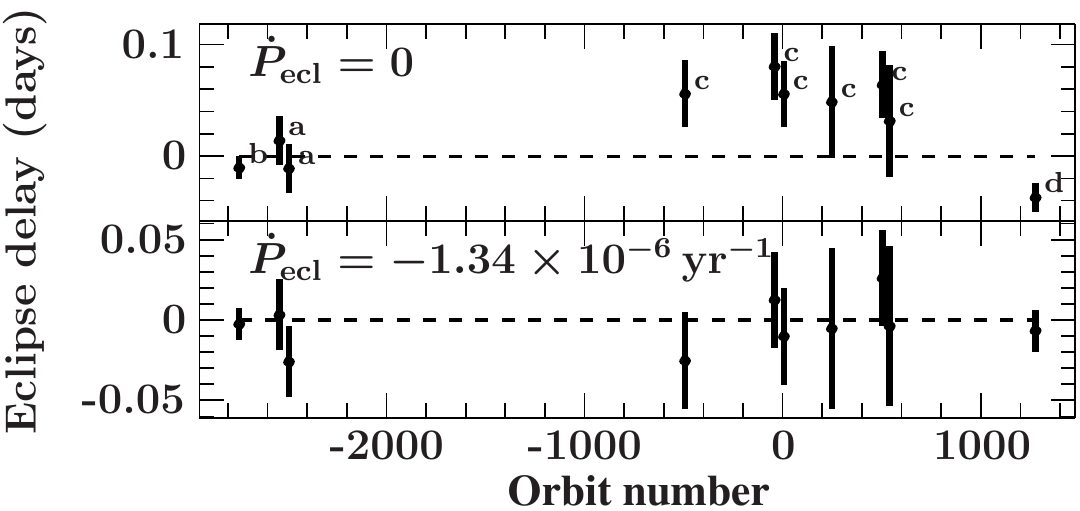}
  \caption{The residuals, in days, to the mid-eclipse times for \fif for
  orbital solutions with no orbital period decay (top) and with nonzero orbital
  period decay (bottom). Measurements are labeled by reference: (a) the \textit{Uhuru} and
  \textit{HEAO-1} measurements of \citet{cominsky_multimission_1991}, (b)
  \textit{Ariel-V}, \citet{davison_binary_1977}, (c) \rxte and \integral,
  \citet{falanga_ephemeris_2015}, and (d) \nustar, this work. The residuals
  without the \textit{Uhuru} measurement are not significantly different.}
  \label{fig:orb}
\end{figure}

Our measured mid-eclipse time is approximately 0.1\,d earlier than the
ephemeris of \citet{falanga_ephemeris_2015} predicts, significantly more than
our uncertainty allows. Thus, we re-fit the mid-eclipse times from
\citet{cominsky_multimission_1991}, \citet{davison_binary_1977}, and
\citet{falanga_ephemeris_2015} along with this new measurement in order to
update the orbital ephemeris. We exclude \citet{davison_binary_1977}'s second
\textit{Ariel-V} measurement and \citet{cominsky_multimission_1991}'s
\textit{EXOSAT} measurements, as these were estimated based on the eclipse
ingress time rather than observing the full eclipse. For an orbital period
measured from mid-eclipse times $P_{\rm ecl}$ and a rate of change in this
period $\dot{P}_{\rm ecl}$, $T_{\rm ecl,n}$, the mid-eclipse time $n$ orbits
after some reference eclipse at $T_{\rm ecl,0}$, is given by
\begin{equation}
  T_{\rm ecl,n} = T_{\rm ecl,0} + nP_{\rm ecl} + \frac{1}{2}P_{\rm ecl}\dot{P}_{\rm ecl}n^2.
  \label{eqn:tn}
\end{equation}
A constant $P_{\rm ecl}$ does not fit the data well, with a $\chi^2$ of
32 for 9 degrees of freedom. Allowing $\dot{P}_{\rm ecl}$ to vary, we
find $\dot{P}_{\rm ecl}/P_{\rm ecl} = \left(-1.27 \pm 0.25\right)
\times 10^{-6}$\,yr$^{-1}$, and the $\chi^2$ improves to 8.02 for 8 degrees of
freedom. Figure~\ref{fig:orb} compares the mid-eclipse delay for
orbital solutions with zero $\dot{P}_{\rm ecl}$ and with $\dot{P}_{\rm
ecl}$ allowed to vary. The \textit{Uhuru} measurement of
\citet{cominsky_multimission_1991} is a $\sim$2$\sigma$ outlier; this may be
simply due to random chance, or it could be related to the observing cadence
and short exposures in the \textit{Uhuru} data (\textit{Uhuru} only observed
the source for approximately two seconds every twelve minutes) combined with
the source's high pulsed fraction. If the \textit{Uhuru} measurement is
excluded, the fit with $\dot{P}_{\rm ecl} = 0$ is still quite poor, with
$\chi^2/{\rm dof} = 28/8$, but the fit with nonzero $\dot{P}_{\rm ecl}$ is
overfitted, with $\chi^2/{\rm dof} = 3.1/7$. However, this does not change the
measured $\dot{P}_{\rm ecl}$ significantly (it increases to $\left(-1.37 \pm
0.25\right) \times 10^{-6}$\,yr$^{-1}$). In the following, we address both the
case with the \textit{Uhuru} measurement and without.

If \fif's orbit is circular, $P_{\rm orb} = P_{\rm ecl}$, and the rate of
change in the eclipse-to-eclipse period is entirely due to orbital period
decay. But for an elliptical orbit, the time between successive eclipses is
not the same as the true orbital period, due to \fif's nonzero apsidal advance
\citep{falanga_ephemeris_2015}. It is relatively simple to calculate $P_{\rm
orb}$ given $P_{\rm ecl}$, $\omega$, $\dot{\omega}$, and the eccentricity $e$.
Following \citet{deeter_apsidal_1987}, but including the higher-order
corrections from their Equation 2, we have
\begin{equation} \label{eqn:porb}
  P_{\rm ecl} - P_{\rm orb} = \frac{eP_{\rm orb}^2}{\pi} f\left(i,\beta\right) \dot{\omega}\sin\omega, \\ 
\end{equation}
where
\begin{align}
  f\left(i,\beta\right) &= \left[1+\frac{1}{2}\cot^2i + \frac{\left(\sin i-\beta\right)\left(1-\beta\sin i\right)}{2\beta\sin^2i}\right] \\
  \beta &= \left[1 - \frac{\left(R/a\right)^2}{1-e^2}\right]^{1/2},
\end{align}
and where we have assumed that $\mathrm{d}f/\mathrm{d}t$ is sufficiently small
that it can be ignored. This is probably a reasonable assumption --- assuming a
change in the semimajor axis of similar order to $\dot{P}_{\rm orb}/P_{\rm
orb}$ as found below, the fractional change in $f$ is of order
$10^{-11}$\,yr$^{-1}$, significantly smaller than our measured $\dot{P}_{\rm
orb}/P_{\rm orb}$.

We then solve the above quadratic equation for $P_{\rm orb}$, finding $P_{\rm
orb} = 3.72831(2)$\,days. Differentiating with respect to time, and assuming
that $\ddot{\omega} = 0$ and $\dot{f} = 0$, we find $\dot{P}_{\rm orb}/P_{\rm
orb} = \left(-1.03 \pm 0.37\right)\times 10^{-6}$\,yr$^{-1}$ when the
\textit{Uhuru} eclipse is excluded, and $\dot{P}_{\rm orb}/P_{\rm orb} =
\left(-0.95 \pm 0.37\right)\times 10^{-6}$\,yr$^{-1}$ if the \textit{Uhuru}
eclipse is included. This is of somewhat marginal significance, although at
$\sim$2.6$\sigma$ it is more significant than previous measurements, most of
which were consistent with zero. We additionally calculated an updated value of
$T_{\pi/2}$ based on \citet{falanga_ephemeris_2015}'s measurements and our new
$P_{\rm ecl}$ and $\dot{P}_{\rm ecl}$ and extrapolated a new value of $\omega$,
the argument of periastron, for this epoch based on
\citet{falanga_ephemeris_2015}'s estimate of an advance of periastron of $1.3
\pm 0.6$\,deg\,yr$^{-1}$. The new orbital ephemeris is displayed in
Table~\ref{tab:orb}.

\subsection{Pulse period and pulsed fraction} \label{sec:pulse}

After correcting for binary motion using our updated ephemeris, we obtained a
new estimate of the pulse period using epoch folding \citep{leahy_searches_1983}, taking only the
out-of-eclipse portions of the lightcurve, as the pulse period is not detected
significantly in the eclipse. This estimate was then refined by folding
segments 0--4 and 13--14 of the lightcurve individually and cross-correlating
their pulse profiles with the average pulse profile to obtain phase shifts,
which, to first order, are linearly related in time to a correction to the
assumed pulse frequency via
\begin{equation}
  \delta\varphi\left(t\right) = \varphi_0 + \delta\nu\left(t-t_0\right),
\end{equation}
where $t_0$ is some reference time (e.g., the start of the observation),
$\varphi_0$ is the phase at this time, $\delta\varphi$ is the measured phase
shift, and $\delta\nu$ is the correction to the assumed frequency $\nu$.
Applying the correction and repeating the process, we find a final pulse period
of $526.41 \pm 0.07$\,s. This is consistent with the value of $\sim$526.5\,s
measured by \textit{Fermi}/GBM around this time. Our measured pulse period is
also presented in Table~\ref{tab:orb}.

% vim: set ft=tex
% % WITHOUT BACKGROUND SUBTRACTION
% \begin{deluxetable}{lr}
%   \tabletypesize{\footnotesize}
%   \tablewidth{0pt}
%   \tablecaption{Orbital parameters and pulse period of \fif \label{tab:orb}.
%   We show 1$\sigma$ errors, and uncertainties in last digits are indicated in
%   parentheses. We include the \textit{Uhuru} mid-eclipse time in this table;
%   see text for the case without \textit{Uhuru}.}
%   \tablehead{\colhead{Parameter} & \colhead{Value}}
%   \startdata
%     $P_{\rm ecl}$ (d) & 3.728360(14) \\
%     $\dot{P}_{\rm ecl}/P_{\rm ecl}$ ($10^{-6}$\,yr$^{-1}$) & $-1.19 \pm 0.41$ \\
%     $P_{\rm orb}$ (d) & 3.72831(3) \\
%     $\dot{P}_{\rm orb}/P_{\rm ecl}$ ($10^{-6}$\,yr$^{-1}$) & $-0.90 \pm 0.49$ \\
%     %$T_{\rm ecl}$ (MJD) & $57612.411(11)$ \\
%     $T_{\pi/2}$ (MJD) & $57612.54(5)$ \\
%     $a\sin i$ (lt-s) & $53.1 \pm 1.5$\tablenotemark{*} \\
%     $\omega$ (degrees) & $57 \pm 14$\tablenotemark{\dag} \\
%     $\dot{\omega}$ (deg\,yr$^{-1}$) & $1.3 \pm 0.6$\tablenotemark{*} \\
%     Eccentricity & 0.18(1)\tablenotemark{*} \\
%     \midrule
%     $P_{\rm pulse}$ (s) & $526.42 \pm 0.07$ \\
%   \enddata
%   %\tablenotetext{*}{Epoch of mean longitude $\pi/2$}
%   \tablenotetext{*}{From \citet{mukherjee_orbital_2006} and \citet{falanga_ephemeris_2015}.}
%   \tablenotetext{\dag}{For the indicated $T_{\pi/2}$, extrapolated using
%   \citet{falanga_ephemeris_2015}'s measurement of $\dot{\omega} = 1.3 \pm
%   0.6$\,deg\,yr$^{-1}$.}
% \end{deluxetable}

% WITH BACKGROUND SUBTRACTION
\begin{deluxetable}{lr}
  \tabletypesize{\footnotesize}
  \tablewidth{0pt}
  \tablecaption{Orbital parameters and pulse period of \fif \label{tab:orb}.
  We show 1$\sigma$ errors, and uncertainties in last digits are indicated in
  parentheses. We include the \textit{Uhuru} mid-eclipse time in this table;
  see text for the case without \textit{Uhuru}.}
  \tablehead{\colhead{Parameter} & \colhead{Value}}
  \startdata
    $P_{\rm ecl}$ (d) & 3.728354(9) \\
    $\dot{P}_{\rm ecl}/P_{\rm ecl}$ ($10^{-6}$\,yr$^{-1}$) & $-1.27 \pm 0.25$ \\
    $P_{\rm orb}$ (d) & 3.72831(2) \\
    $\dot{P}_{\rm orb}/P_{\rm ecl}$ ($10^{-6}$\,yr$^{-1}$) & $-0.95 \pm 0.37$ \\
    %$T_{\rm ecl}$ (MJD) & $57612.411(11)$ \\
    $T_{\pi/2}$ (MJD) & $57612.53(5)$ \\
    $a\sin i$ (lt-s) & $53.1 \pm 1.5$\tablenotemark{*} \\
    $\omega$ (degrees) & $57 \pm 14$\tablenotemark{\dag} \\
    $\dot{\omega}$ (deg\,yr$^{-1}$) & $1.3 \pm 0.6$\tablenotemark{*} \\
    Eccentricity & 0.18(1)\tablenotemark{*} \\
    \midrule
    $P_{\rm pulse}$ (s) & $526.42 \pm 0.07$ \\
  \enddata
  %\tablenotetext{*}{Epoch of mean longitude $\pi/2$}
  \tablenotetext{*}{From \citet{mukherjee_orbital_2006} and \citet{falanga_ephemeris_2015}.}
  \tablenotetext{\dag}{For the indicated $T_{\pi/2}$, extrapolated using
  \citet{falanga_ephemeris_2015}'s measurement of $\dot{\omega} = 1.3 \pm
  0.6$\,deg\,yr$^{-1}$.}
\end{deluxetable}

\begin{figure}
  \centering
  \plotone{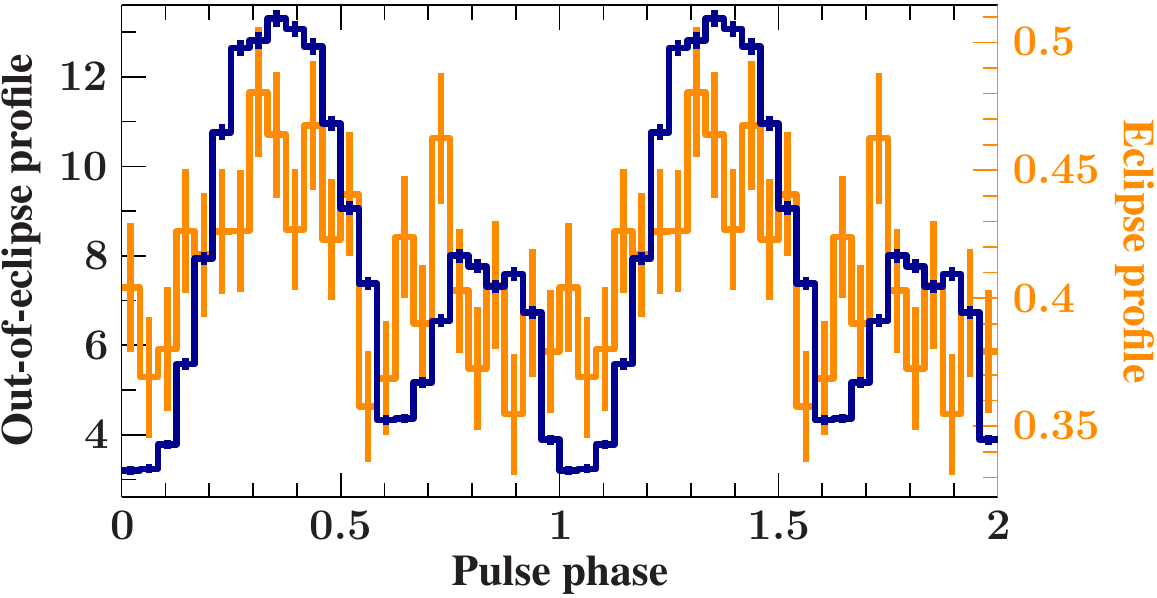}
  \caption{The pulse profiles of the pre-eclipse lightcurve (blue) and eclipse
  (gold).}
  \label{fig:prof}
\end{figure}

Figure~\ref{fig:prof} displays the pulse profile of the pre-eclipse lightcurve
after folding on the measured pulse period. Epoch folding does not detect
significant pulsations during eclipse, and while the eclipse profile (also in
Figure~\ref{fig:prof}), when folded on the pre-eclipse pulse period, does show
weak pulsations, the pulsed fraction is very low. We define the pulsed
fraction $PF$ in the usual manner:
\begin{equation}
  PF = \frac{F_{\rm max} - F_{\rm min}}{F_{\rm max} + F_{\rm min}},
\end{equation}
where we define the maximum flux $F_{\rm max}$ and minimum flux $F_{\rm min}$
as the average flux from the three highest- and lowest-flux points in a pulse
profile in order to average over variations during pulse maxima and minima. We
propagate the 1$\sigma$ errors in the counting rates to determine the
uncertainty in the pulsed fraction.

\begin{figure*}
  \centering
  \plottwo{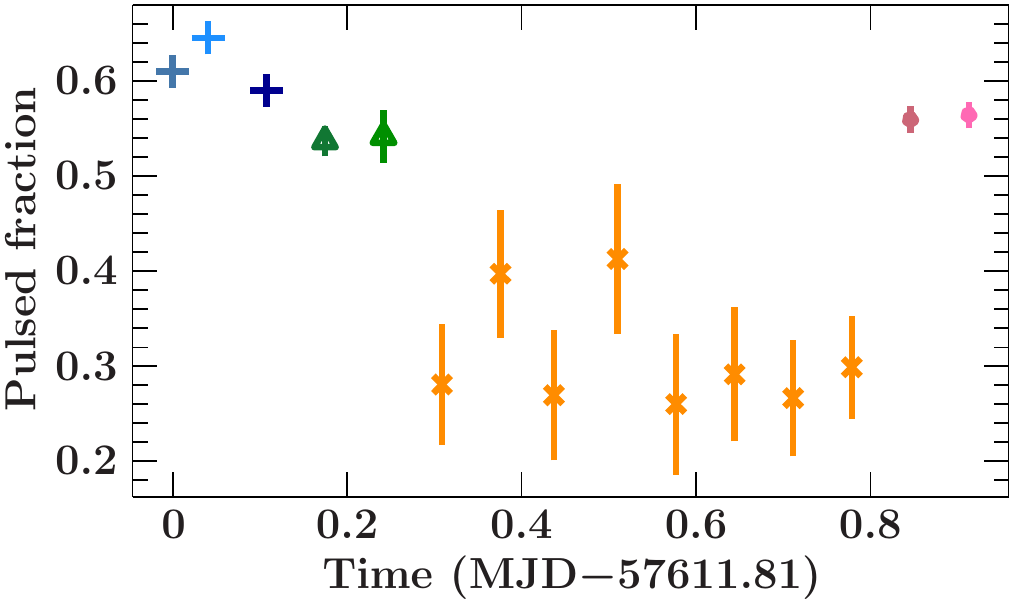}{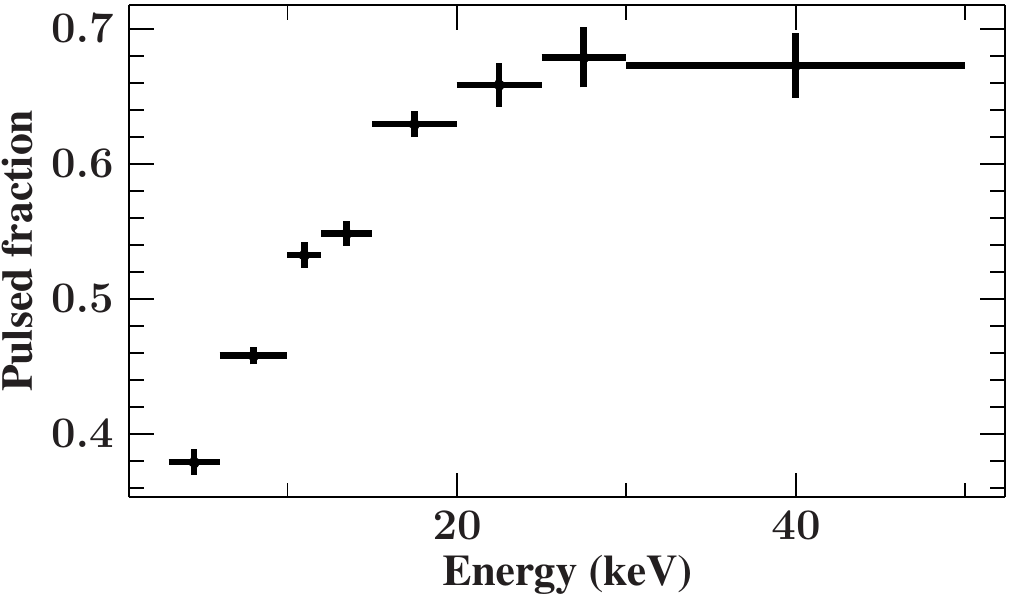}
  \caption{Left: the pulsed fraction in the full 3--78\,keV lightcurve as a
  function of time, with colors and symbols same as in Figure~\ref{fig:lc}.
  Right: the pulsed fraction in the out-of-eclipse phases of the observation,
  plotted as a function of energy.}
  \label{fig:pulsedfraction}
\end{figure*}

As displayed in the left panel of Figure~\ref{fig:pulsedfraction}, during
eclipse the pulsed fraction ranges from 0.2 to 0.4, compared to values of
0.5--0.65 for the out-of-eclipse phases. The out-of-eclipse lightcurve also
shows considerable variation in pulsed fraction with energy, as can be seen in
the right-hand panel of Figure~\ref{fig:pulsedfraction}, showing a rise from
$\sim$0.4\ in the 3--6\,keV band to $\sim$0.7\ above 20\,keV, flattening off
after that.

\section{Spectral analysis} \label{sec:spec}

As noted in Section~\ref{sec:obs}, we extracted spectra for the pre-eclipse
(10.5\,ks exposure), ingress (4.4\,ks), eclipse (25.1\,ks), and egress
(6.2\,ks). Spectra were rebinned to a minimum combined FPMA+FPMB signal-to-noise
of 6. All fits were carried out with a lower bound of 3\,keV and an
upper bound determined by the quality of the data (50, 45, 35, and 50\,keV for
the pre-eclipse, ingress, eclipse, and egress spectra, respectively). In all
cases we fit the FPMA and FPMB spectra simultaneously, with a
cross-normalization constant ($C_{\rm FPMB}$) to account for minor differences
in area between the two telescopes (although in plots, for clarity, we show the
combined FPMA+FPMB spectra). For completeness, we fit these spectra with
several different empirical continuum models. First, as in
\citet{hemphill_evidence_2016}, we employ an empirical continuum consisting of
a power law in energy multiplied by a high-energy cutoff above some energy
$E_{\rm cut}$ \citep{white_accretion_1983}:
\begin{eqnarray}
  \mathtt{plcut}(E) &=& \left\{\begin{array}{cc}
    A E^{-\Gamma} & E < E_{\mathrm{cut}} \\
    A E^{-\Gamma}\exp\left( \frac{E_{\mathrm{cut}} - E}{E_{\mathrm{fold}}} \right) & E \ge E_{\mathrm{cut}}.
\end{array}\right.
  \label{eqn:plcut}
\end{eqnarray}
As this piecewise continuum model is known to introduce absorption-like
residuals around the cutoff energy $E_{\rm cut}$, we include a narrow negative
Gaussian with its energy tied to the cutoff energy to ``smooth'' the
introduction of the cutoff, as used by, e.g., \citet{coburn_study_2001}. We
refer to this ``modified'' cutoff power law as \texttt{mplcut}. We also carried
out fits using other continuum models: a power law with the ``Fermi-Dirac''
high-energy cutoff model from \citet{tanaka_observations_1986}:
\begin{equation}
  \mathtt{fdco}(E) = A E^{-\Gamma} \times \left( 1 + \exp\left[ \frac{E - E_{\rm cut}}{E_{\rm fold}} \right] \right)^{-1},
  \label{eqn:fdcut}
\end{equation}
and the Negative-Positive EXponential ``\texttt{npex}'' continuum from
\citet{mihara_observational_1995}:
\begin{equation}
  \mathtt{npex}(E) = A \times \left( E^{-\alpha} + A_{\rm p}E^{-\beta} \right)\times \exp\left( -\frac{E}{E_{\rm fold}} \right).
  \label{eqn:npex}
\end{equation}
In the \texttt{npex} model, the $\alpha$ parameter is positive-definite and the
$\beta$ parameter is negative-definite, creating a model with one
negative-index power law and one positive-index power law. Note that in our
implementation, the $A_{\rm p}$ parameter is the normalization of the positive
power law relative to the negative power law.

As in \citet{hemphill_evidence_2016}, the fundamental and harmonic CRSFs are
modeled with multiplicative Gaussian optical depth profiles, \texttt{gauabs}:
\begin{eqnarray}
  \mathtt{gauabs}(E) &=& e^{-\tau\left(E\right)} \\
  \tau(E) &=& \tau_{0}\exp\left( -\frac{\left(E-E_{0}\right)^{2}}{2\sigma^{2}}\right).
  \label{eqn:gauabs}
\end{eqnarray}
Note that \texttt{gauabs} has the same mathematical form as the \texttt{gabs}
model, with the exception of how the depth of the line is defined. We include
the harmonic CRSF only in the pre-eclipse spectrum, as the other spectra do not
extend to high enough energies. We also fix its energy to 50\,keV, based on
\citet{rodes-roca_first_2009} and \citet{hemphill_measurements_2013}, as it was
poorly constrained and at the upper edge of the useful \nustar data. The width
of the harmonic CRSF could only be constrained when using the \texttt{mplcut}
continuum, so we freeze its value to the \texttt{mplcut}-derived value
(10\,keV) in all other fits.

The source is highly absorbed during this observation, even out of eclipse, as
it is passing behind the limb of \qvnor. We applied the \texttt{TBabs}
absorption model\footnote{See
\url{http://pulsar.sternwarte.uni-erlangen.de/wilms/research/tbabs/}}
\citep{wilms_dont_2010}, with cross sections from \citet{verner_atomic_1996}
and ISM abundances from \citet{wilms_absorption_2000}, to handle this
low-energy absorption. We employ the \texttt{enflux} spectral component in ISIS
to compute the unabsorbed model flux in the 3--50\,keV band. The iron line
complex shows a prominent K$\alpha$ line with some additional structure in the
residuals around 7\,keV, which we model with Gaussians. The pre-eclipse
residuals are fit well by a pair of narrow lines at energies consistent with
neutral Fe K$\alpha$ and K$\beta$, while the higher contrast of the eclipse
spectrum shows more complexity and requires three lines (see
Section~\ref{sec:ecl}).

The \rxte spectra analyzed in \citet{hemphill_evidence_2016} required
additional absorption around 8\,keV (or possibly emission around 12\,keV) to
obtain a good fit. The \nustar data do not require this feature (the
statistical uncertainties on our spectral points are larger than the depth of
the feature measured by \rxte), but to best compare to the results of
\citet{hemphill_evidence_2016}, we verified that including a third
\texttt{gauabs} feature with its width fixed to 1\,keV as in
\citet{hemphill_evidence_2016} did not change our results significantly.

The final models used for the pre-eclipse spectrum are thus of the form
\begin{eqnarray}
  \mathtt{TBabs} &\times& (\mathtt{enflux}(\mathtt{<continuum>} \nonumber \\
                 &\times& \mathtt{gauabs_{\rm fund.}} \times \mathtt{gauabs_{\rm harm.}}) \nonumber \\
&+& \mathtt{gauss}_{6.4 \rm keV} + \mathtt{gauss}_{7 \rm keV})
\end{eqnarray}
where \texttt{<continuum>} is Equation~\ref{eqn:plcut} (\texttt{mplcut}, in
which case there is a third, negative \texttt{gauss} component to ``smooth'' the
\texttt{highecut} model), \ref{eqn:fdcut} (\texttt{fdco}), or \ref{eqn:npex}
(\texttt{npex}). In the eclipse ingress, eclipse, and eclipse egress spectra we
add a \texttt{TBpcf} component to model a partial-covering absorber (see
Section~\ref{sec:ecl}) and omit the second \texttt{gauabs} feature.

\subsection{Pre-eclipse spectral fits} \label{sec:pre_results}

\begin{figure}
  \centering
  \plotone{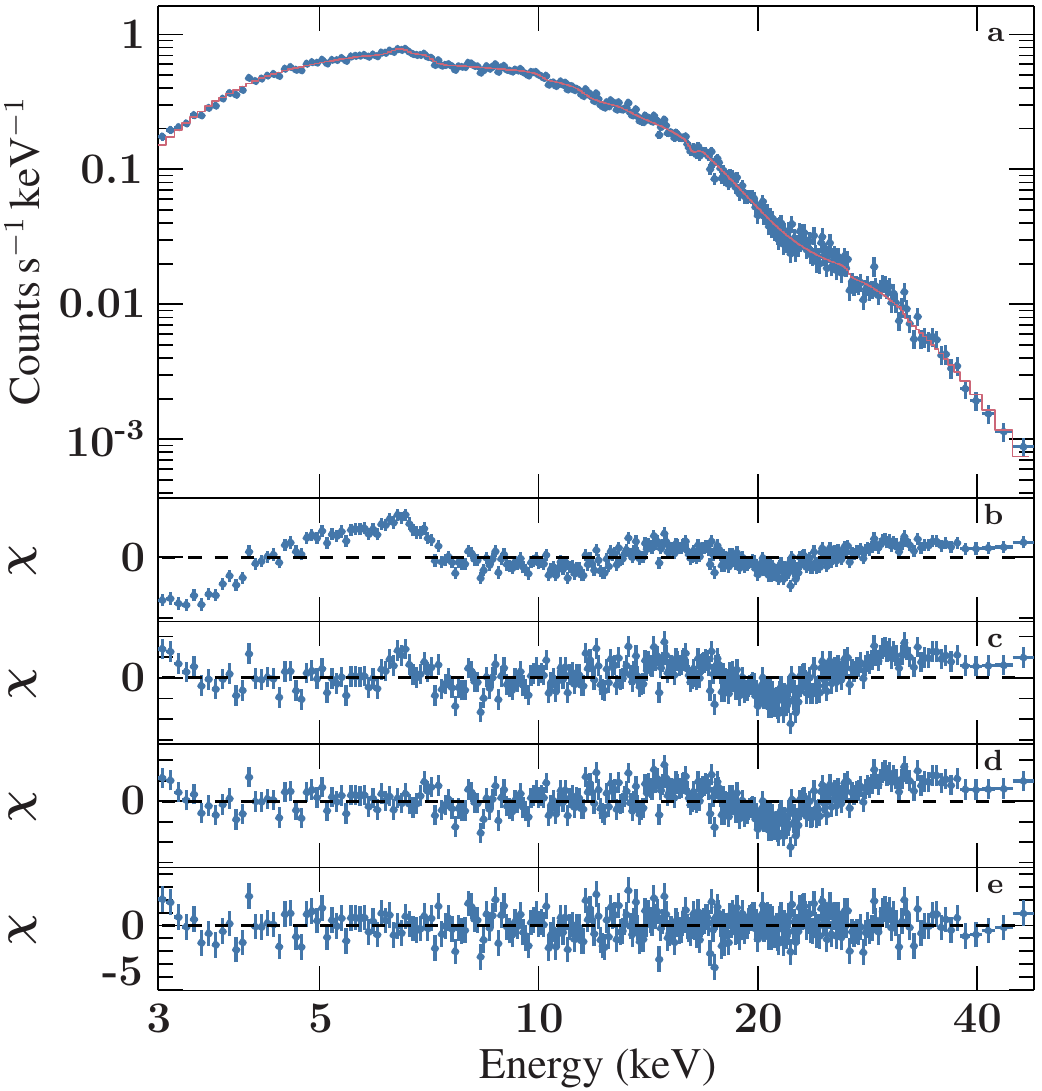}
  \caption{The pre-eclipse spectrum of \fif, fitted with the \texttt{mplcut}
    continuum model. The two \nustar detectors have been added for clarity.
    Panel a displays the spectrum with the best-fit model overplotted.  Further
    panels show residuals to fits starting with only a
    \texttt{powerlaw*highecut} continuum (panel b), adding in absorption (panel
    c), iron emission lines (panel d), and finally the CRSFs (panel e).}
  \label{fig:pre}
\end{figure}

The best-fitting parameters for the pre-eclipse spectrum and their 90\% error
bars are displayed in Table~\ref{tab:pre}. The combined FPMA+FPMB spectrum with
the best-fit \texttt{mplcut} model is plotted in Figure~\ref{fig:pre}. The three
continuum models find similarly good fits.

\begin{deluxetable*}{lhrhhrr}
  \tabletypesize{\footnotesize}
  \tablewidth{0pt}
  \tablecaption{Pre-eclipse best-fit parameters for \fif\label{tab:pre}}
  \tablehead{\colhead{}&\nocolhead{\texttt{plcut}}&\colhead{\texttt{mplcut}}&\nocolhead{\texttt{mplcut}, 8\,keV dip}&\nocolhead{\texttt{mplcut}, minus eclipse}&\colhead{\texttt{fdco}}&\colhead{\texttt{npex}}}
  \startdata
  Flux\tablenotemark{a} ($10^{-10}$\,\eflx) & $6.42\pm0.10$ & $6.44^{+0.10}_{-0.08}$ & $6.46\pm0.10$ & $6.22^{+0.10}_{-0.08}$ & $6.31\pm0.10$ & $6.6^{+0.3}_{-0.2}$ \\ 
    $C_{\rm FPMB}$ & $1.026\pm0.011$ & $1.025\pm0.011$ & $1.026\pm0.011$ & $1.025^{+0.012}_{-0.011}$ & $1.025\pm0.011$ & $1.026\pm0.011$ \\ 
    \midrule
    $N_{\rm H}$ ($10^{22}$\,cm$^{-2}$) & $14.8^{+1.1}_{-1.0}$ & $14.7\pm1.1$ & $15.1\pm1.1$ & $15.3\pm1.1$ & $13.5^{+1.1}_{-1.2}$ & $17.6^{+3.1}_{-2.7}$ \\ 
    $\Gamma$ & $1.19\pm0.04$ & $1.19 \pm 0.04$ & $1.19^{+0.05}_{-0.07}$ & $1.21\pm0.05$ & $1.06^{+0.06}_{-0.07}$ & \nodata \\ 
    $E_{\rm cut}$ (keV) & $14.6^{+0.7}_{-0.5}$ & $16.27^{+0.14}_{-2.09}$ & $16.25^{+0.16}_{-0.14}$ & $16.25^{+0.15}_{-0.12}$ & $24.4^{+1.6}_{-1.9}$ & \nodata \\ 
    $E_{\rm fold}$ (keV) & $14.0^{+3.7}_{-1.4}$ & $12.4^{+5.7}_{-0.6}$ & $12.9^{+7.6}_{-1.2}$ & $12.4^{+1.5}_{-1.2}$ & $6.3\pm0.6$ & $4.7^{+1.0}_{-0.8}$ \\ 
    Smoothing area ($10^{-4}$\,\pflx) & \nodata & $0.7\pm0.4$ & $< 0.943$ & $0.7\pm0.4$ & \nodata & \nodata \\ 
    $\alpha$ & \nodata & \nodata & \nodata & \nodata & \nodata & $0.90^{+0.79}_{-0.28}$ \\ 
    $\beta$ & \nodata & \nodata & \nodata & \nodata & \nodata & $-2.1^{+0.9}_{-1.1}$ \\ 
    $A_{\rm p}$ ($10^{-4}$\,\pflx) & \nodata & \nodata & \nodata & \nodata & \nodata & $16^{+30}_{-13}$ \\ 
    % $E_{\rm dip}$ (keV) & $(8.38)$ & $(8.38)$ & $8.4\pm0.8$ & $8.2\pm0.9$ & $(8.62)$ & \nodata \\ 
    % $\tau_{\rm dip}$ (keV) & $(0.000)$ & $(0.000)$ & $0.039^{+0.020}_{-0.019}$ & $0.036\pm0.020$ & $(0.000)$ & \nodata \\ 
    \midrule
    $E_{\rm fund}$ (keV) & $22.3\pm0.4$ & $22.0\pm0.4$ & $22.1\pm0.4$ & $21.9\pm0.4$ & $22.81^{+0.29}_{-0.28}$ & $22.53^{+0.28}_{-0.27}$ \\ 
    $\sigma_{\rm fund}$ (keV) & $2.9\pm0.4$ & $3.18^{+0.28}_{-0.27}$ & $3.3^{+0.4}_{-0.0}$ & $3.3^{+0.4}_{-0.0}$ & $3.29^{+0.26}_{-0.25}$ & $3.4^{+0.5}_{-0.4}$ \\ 
    $\tau_{\rm fund}$ (keV) & $0.52^{+0.09}_{-0.07}$ & $0.60^{+0.07}_{-0.06}$ & $0.61^{+0.09}_{-0.08}$ & $0.59^{+0.06}_{-0.00}$ & $0.78\pm0.07$ & $0.74^{+0.11}_{-0.08}$ \\ 
    $E_{\rm harm}$ (keV) & $(50.0)$\tablenotemark{b} & $(50.0)$ & $(50.0)$ & $(50.0)$ & $(50.0)$ & $(50.0)$ \\ 
    $\sigma_{\rm harm}$ (keV) & $9\pm4$ & $9\pm5$ & $10\pm6$ & $(10.0)$ & $(10.0)$ & $(10.0)$ \\ 
    $\tau_{\rm harm}$ (keV) & $0.5^{+0.6}_{-0.4}$ & $0.4^{+0.6}_{-0.4}$ & $0.49^{+0.29}_{-0.42}$ & $< 0.802$ & $< 0.134$ & $< 0.180$ \\ 
    \midrule
    Fe K$\alpha$\tablenotemark{c} area ($10^{-4}$\,\pflx) & $1.7\pm0.6$ & $1.9^{+0.6}_{-0.5}$ & $1.80^{+0.48}_{-0.14}$ & $0.7\pm0.6$ & $2.0^{+0.5}_{-0.6}$ & $2.1\pm0.6$ \\ 
    Fe K$\alpha$ energy (keV) & $(6.40)$ & $6.44\pm0.07$ & $6.45^{+0.07}_{-0.06}$ & $6.44^{+0.20}_{-0.25}$ & $6.46^{+0.04}_{-0.09}$ & $6.45^{+0.07}_{-0.06}$ \\ 
    7\,keV line area ($10^{-4}$\,\pflx) & $0.5\pm0.6$ & $0.7\pm0.5$ & $0.6^{+0.6}_{-0.5}$ & $0.32873560^{+0.00000011}_{-0.00000160}$ & $0.6\pm0.5$ & $0.8\pm0.5$ \\ 
    7\,keV line energy (keV) & $(6.68)$ & $6.96^{+0.13}_{-0.17}$ & $6.97^{+0.16}_{-0.21}$ & $(6.98)$ & $(6.98)$ & $6.98^{+0.15}_{-0.14}$ \\ 
    % Fe K$\beta$ area ($10^{-4}$\,\pflx) & $0.5\pm0.5$ & \nodata & \nodata & \nodata & \nodata & \nodata \\ 
    % Fe K$\beta$ energy (keV) & $(7.06)$ & \nodata & \nodata & \nodata & \nodata & \nodata \\ 
    \midrule
    $\chi^{2}_{\rm red}$ & 0.99 (632) & 0.99 (630) & 0.98 (628) & 0.97 (608) & 1.08 (633) & 1.01 (631) \\ 
  \enddata
  \tablenotetext{a}{Unabsorbed flux in 3--50\,keV band.}
  \tablenotetext{b}{Values in parentheses were frozen during fitting.}
  \tablenotetext{c}{K$\alpha$ and 7\,keV line widths fixed to 0.01\,keV.}
\end{deluxetable*}

It is apparent that the source was at a relatively low luminosity during the
observation, even when the absorption is taken into account. The unabsorbed 3--50\,keV flux
of $6.6\times10^{-10}$\,\eflx, which translates to a
luminosity in the same band of $3.2\times10^{36}$\,erg\,s$^{-1}$ at a distance of $6.4$\,kpc, is
comparable to the lowest-flux \rxte observations
\citep[see][]{hemphill_evidence_2016}. This likely still places \fif in the
super-critical (radiation-dominated) regime, as the critical luminosity in this
source is $\sim\!10^{36}$\,erg\,s$^{-1}$ \citep{hemphill_evidence_2016}. The
continuum parameters measured using the \texttt{mplcut} model are consistent
with the overall picture of the source as seen by \rxte, which at these fluxes
found a photon index of $\sim\!1.1$, a cutoff energy near $15$\,keV, and a
folding energy of $\sim\!12$\,keV. The iron line complex in \nustar is best fit with a
pair of narrow lines around 6.4 and 7\,keV, consistent with the iron K$\alpha$
and K$\beta$ features. However, the flux of the 7\,keV line is $\sim$37\% of
the K$\alpha$ flux, quite high for a K$\beta$ line. This may be due to a strong
iron K edge, or due to blending with higher-ionization iron emission lines.
Based on the eclipse spectrum from this observation and the \chandra gratings
results \citep{torrejon_radial_2015}, this is most likely a blend of
\ion{Fe}{25} and the neutral iron K$\beta$, although formally we cannot
rule out contributions from \ion{Fe}{26}.

While the majority of the spectral parameters are consistent with the average
\rxte parameters, the energy of the fundamental CRSF is more comparable to that
found by \suz, at $22.0 \pm 0.4$\,keV, consistent with
\citet{hemphill_evidence_2016}'s conclusion that the CRSF energy had shifted
upwards between the \rxte and \suz measurements. Our fits with the \texttt{fdco}
and \texttt{npex} continua also find values consistent with those reported in \citet{hemphill_clumpy_2014}
using \suz. We can estimate the probability of this arising by chance in
multiple ways. First, fitting a constant $E_{\rm cyc}$ to the
observation-by-observation \rxte, \suz, and \nustar CRSF energy measurements
finds a poor $\chi^2$ of 171.3 for 51 degrees of freedom. However, a considerable portion of this is due to the scatter in the \rxte
measurements --- the constant-$E_{\rm cyc}$ fit to the \rxte data alone finds a
$\chi^2$ of 128.9 with 49 degrees of freedom. To take the scatter of the \rxte
points into account, we take a similar approach as was used in
\cite{hemphill_evidence_2016}. We assume that the set of obsid-by-obsid \rxte
measurements and errors ($N = 50$) can be used as a baseline distribution of
CRSF measurements for this source --- i.e., our null hypothesis is that the
\nustar measurement comes from the same distribution as the \rxte measurements.
We then produce $10^{7}$ simulated sets of CRSF energy measurements by randomly
scattering our \nustar measurement and the 50 obsid-by-obsid \rxte measurements
from \citet{hemphill_evidence_2016} according to their measured 1$\sigma$
uncertainties. For each trial, we compute the distance in standard deviations
between the simulated \nustar measurement and the simulated \rxte points. A
Gaussian fit to the resulting distribution of distances finds the \nustar point
$3.1\sigma$ above the \rxte measurements. In order to approximate the
significance of the \suz and \nustar measurements together, we note that out of
the $10^7$ simulated \rxte datasets, 0.03\% have multiple measurements higher
than the \suz and \nustar CRSF energies. This corresponds to a significance of
approximately $3.6\sigma$. When the 21.9\,keV CRSF measured by
\textit{AstroSat} is taken into account, the statistical significance increases
to $\sim$4$\sigma$. We therefore conclude that, at least based on statistical
errors, it is unlikely for the higher energies measured by \suz and \nustar to
be purely by chance.

\begin{figure}
  \centering
  \plotone{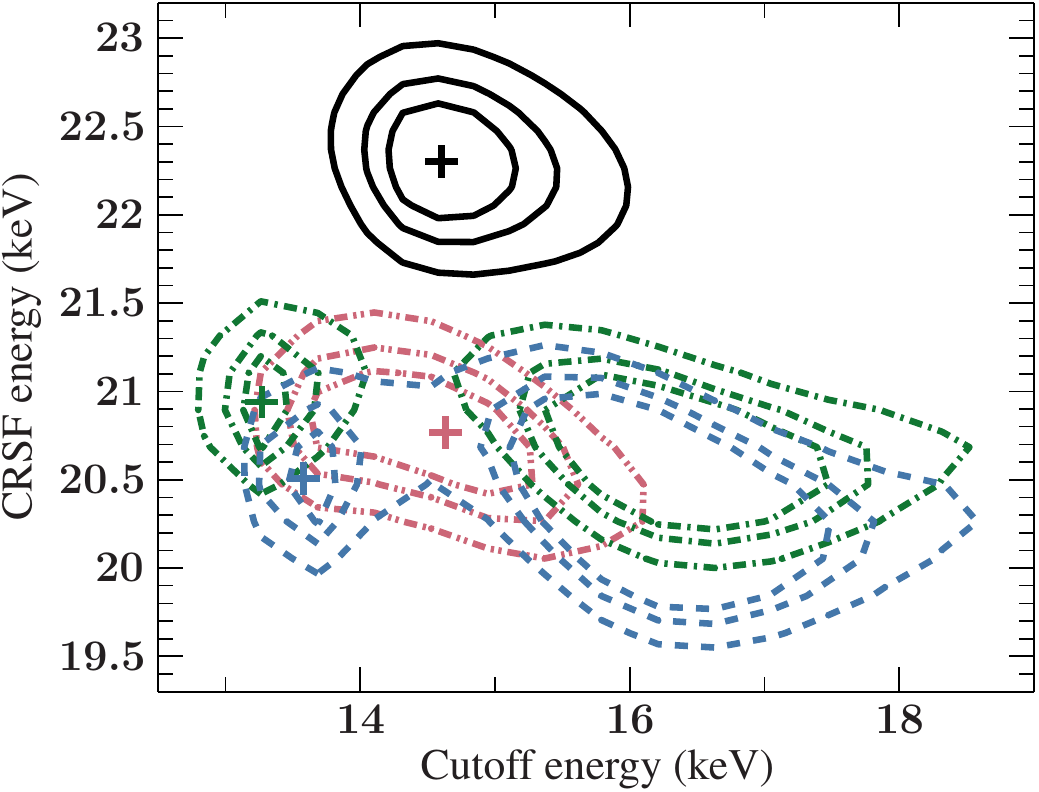}
  \caption{$E_{\rm cyc}$ vs. $E_{\rm cut}$ contours for \nustar (solid black lines) and \rxte
  from proposals P10145 (blue dashed), P50067 (green dot-dashed), and P80016
  (red dot-dot-dashed). The best-fit values for each set of contours are plotted
  with crosses. The contours represent, moving outward from the center,
  $\Delta\chi^{2}$ values of 2.3, 4.61, and 9.21, corresponding to the 68\%,
  90\%, and 99\% confidence limits.}
  \label{fig:contours}
\end{figure}

As an attempt to rule out systematic differences between the \rxte and \nustar
results, we computed two-dimensional confidence contours between the cyclotron
line energy and the cutoff energy for the \texttt{mplcut} continuum, as the
piecewise \texttt{highecut} model can produce spurious absorption-like
residuals around the cutoff energy which can interfere with the CRSF
measurement. The \nustar confidence contours are displayed in
Figure~\ref{fig:contours}, overplotted with \rxte contours from the low-flux
(PCU2 counting rates between 103 and 123 counts\,s$^{-1}$) phase-averaged
spectra presented in \citet{hemphill_evidence_2016}. The \nustar contours are
very well-behaved despite the low flux, and the \rxte contours are all excluded
at better than the 99\% level. Additionally, unlike the \suz contours, there
are no regions where the CRSF energies overlap even at different values of
$E_{\rm cut}$ \citep[\textit{cf.} Figure 9 in][]{hemphill_evidence_2016}. This
feature in the \suz contours was likely due to the gap in coverage between the
XIS and HXD/PIN spectra and the associated uncertainty in the cross-calibration
constant between those instruments; \nustar's single-instrument coverage of its
entire energy band avoids this issue.

\subsection{Ingress, eclipse, and egress spectra} \label{sec:ecl}

\begin{figure}
  \centering
  \plotone{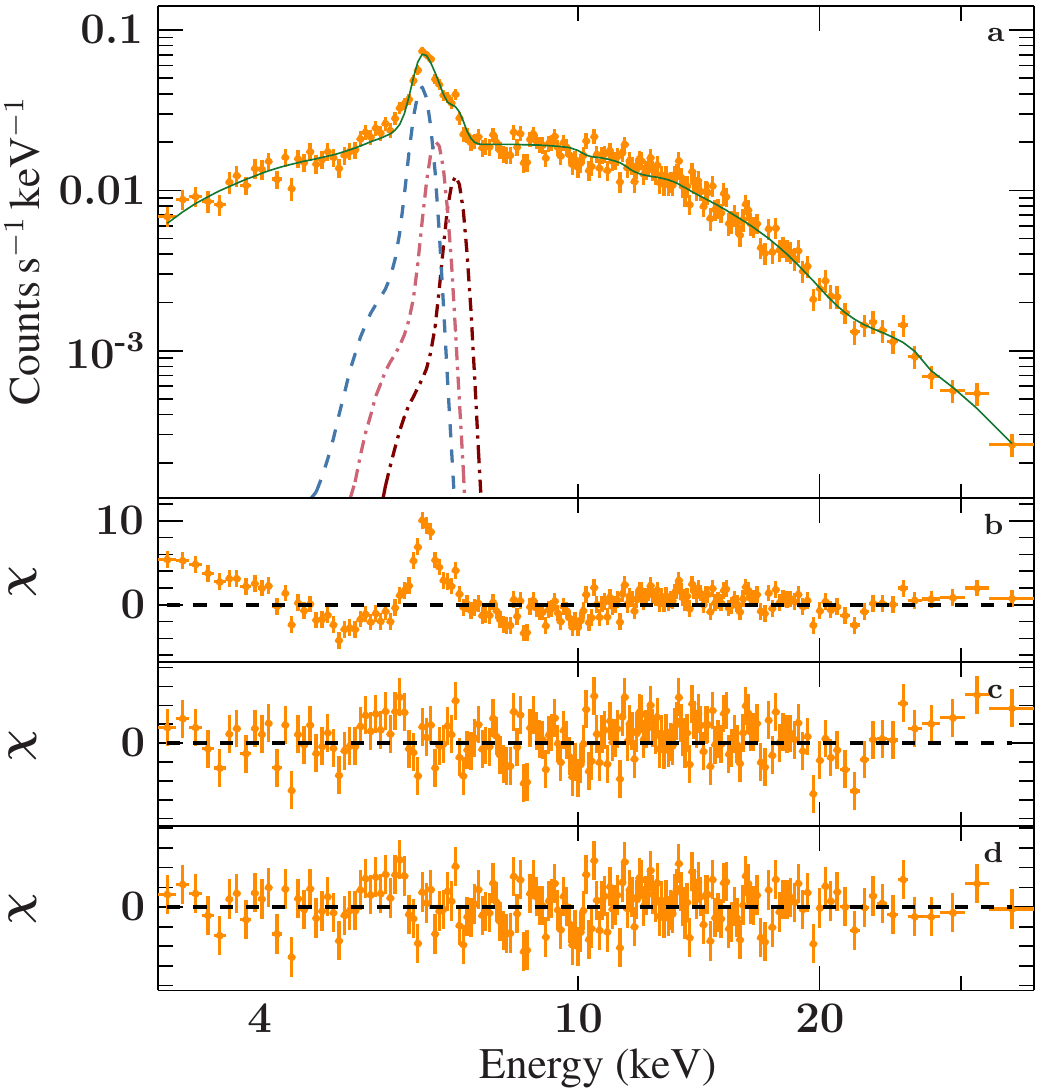}
  \caption{Panel a plots the eclipse spectrum and best-fit model, with the model
    components for the three iron lines features overplotted. Further panels
    display the residuals to fits with no emission lines or CRSF (panel b),
    with the three narrow iron lines (panel c), and finally with a CRSF at
    $21.7$\,keV (panel d).}
  \label{fig:ecl}
\end{figure}

The eclipse spectrum has a total exposure of 25.1\,ks, and is plotted in
Figure~\ref{fig:ecl}. We analyze the individual numbered ingress and egress
spectra as presented in Figure~\ref{fig:spectra}, due to the large spectral
variations during these phases. Spectra 3 and 4 (the eclipse ingress) each have
exposures of 2.4\,ks, while spectra 13 and 14 (the eclipse egress) have
exposures of 3.2 and 3\,ks, respectively. The eclipse profile is similar to
previous observations \citep[see, e.g.,][]{rubin_observation_1997}
and resembles that of the similar wind-fed X-ray pulsar Vela~X-1
\citep{sato_xray_1986}. Our best-fit models for the ingress, eclipse, and
egress spectra are displayed in Table~\ref{tab:pcf}, and the eclipse spectrum
with its best-fit model is displayed in Figure~\ref{fig:ecl}. The model we use
for all these spectra incorporates a partial-covering absorber and three
Gaussian emission lines, as explained below.

The partial-covering absorption model is chosen in order to fit the ingress,
eclipse, and egress consistently. Note that this model is arguably
unphysical for the eclipse spectrum, which is observed only through scattered
emission. The eclipse spectrum, taken alone, is well-fit by
a single, full-covering absorber, but the absorbing column can only be constrained to an upper limit of $\sim
5\times10^{22}$. This is lower than \nh during the pre-eclipse phase and
is consistent with the ISM absorption of
$1\times10^{22}$\,cm$^{-2}$, which in turn is consistent with the pulsar's
direct emission being completely blocked (and only seen through scattered
emission) rather than absorbed. However, the eclipse ingress and egress spectra
are not well-fit by a single absorber, and the measured \nh is strongly
correlated with the power-law index. This is especially the case in spectrum 4,
where the extreme spectral slope below 10\,keV (see Figure~\ref{fig:spectra})
prefers a relatively \textit{low} measured \nh and a highly negative photon
index, which we reject as unphysical. Thus, to investigate the change in
absorption across the eclipse, we fit all three spectra with the fully-covered
\nh fixed to the $14.7\times 10^{22}$\,cm$^{-2}$ column measured in the
pre-eclipse spectrum, with a partial-covering absorption model,
\texttt{tbnew\_pcf}, to model the changing part of the absorbing column.  The
partially-covered \nh, \nhpc, and covering fraction, \fpc, are both left free
to vary.  We also fix the photon index to 1.19, as measured in the pre-eclipse
spectrum --- that is, we assume that the changes in the low-energy spectrum are
entirely due to changes in the partial coverer. Under these prescriptions, we
find that the covering fraction and partially-covered \nh increase as the
source moves into eclipse and decrease as it moves out. \nh appears to be
higher during eclipse ingress compared to egress, consistent with the energy
dependence of the asymmetric eclipse profile seen in Figure~\ref{fig:lc}. In
eclipse, both quantities are poorly constrained; the partially-covered \nh is
consistent with \nh immediately to either side of the eclipse, but the covering
fraction is lower.

The iron line complex also requires more careful treatment than it did in the
pre-eclipse spectrum. In eclipse, the broad-band flux drops by a factor of
$\sim$30 compared to the pre-eclipse phase, while the flux of the iron lines
does not decrease by a measurable amount. This is similar to what was seen using
\textit{Chandra} by \citep{torrejon_radial_2015}, who found a factor of $\sim$3
decrease in the line flux in eclipse compared to a 97\% drop in broad-band
flux. This higher-contrast look at the iron line complex reveals it be
considerably more complex than was seen in the pre-eclipse spectrum. The
residuals are broad and asymmetric, not fit well by a single line. Two
Gaussians with variable energy and width prefer a narrow feature at
$\sim$6.5\,keV and a broader second peak at $\sim$6.6\,keV. However, these line
energies are problematic from a theoretical standpoint --- a narrow iron line
at 6.5\,keV does not line up with any physically reasonable range of
ionizations \citep{kallman_photoionization_2004}. The basic structure in the
residuals that we observe --- one narrow and one broad line --- is
superficially similar to what is seen in \textit{XMM-Newton}, but our lines are
$\sim$0.1\,keV, approximately twice the typical uncertainty we find on the line
energies, above the 6.4 and 6.5\,keV lines reported by
\citet{gimenez-garcia_xmm-newton_2015} using that dataset.

\begin{figure}
  \centering
  \plotone{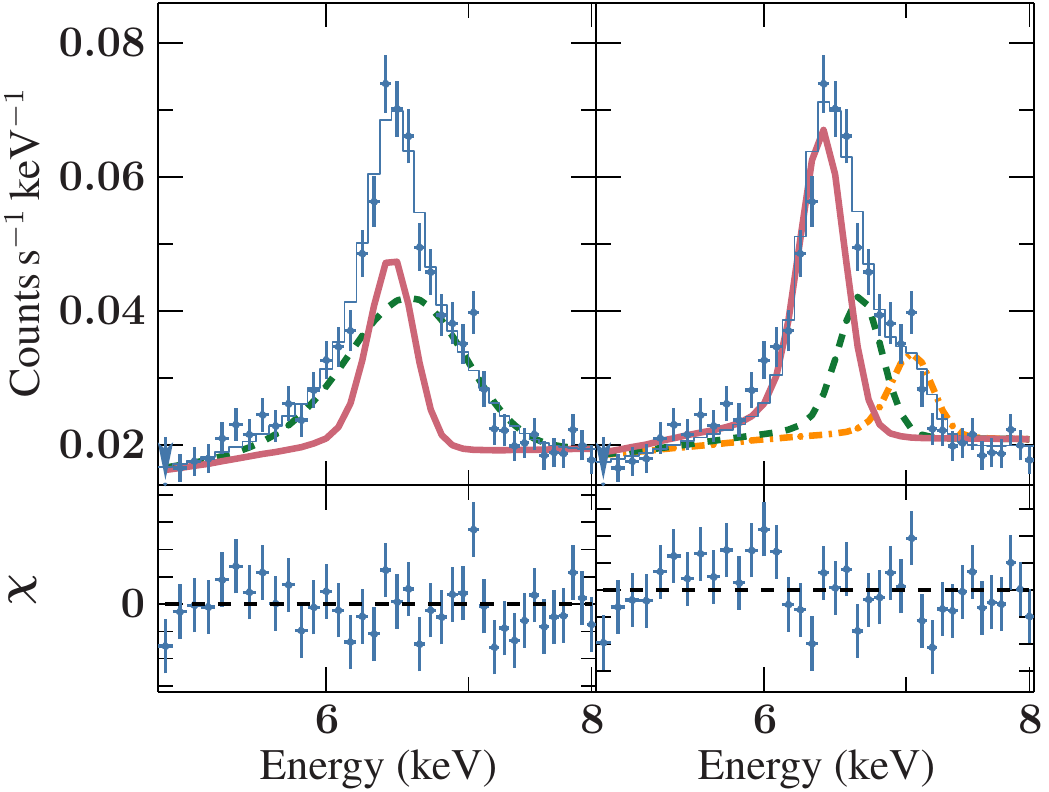}
  \caption{A comparison between the formal best-fit model, with a narrow line
    at $\sim$6.5\,keV and a broad feature at $\sim$6.6\,keV (left) and a fit
    with narrow Fe K$\alpha$, \ion{Fe}{25}, and Fe K$\beta$ as found in
    \textit{Chandra} gratings spectra (right). The blue points and histogram
    are the data and best-fit model, respectively. Fe K$\alpha$ is plotted as a
    solid red line, \ion{Fe}{25} in dashed green, and Fe K$\beta$ in dot-dashed
    gold. The fit with the three narrow lines is moderately worse, underfitting
    the red wing of the iron line, but is more physically reasonable (see text).}
    \label{fig:compare_lines}
\end{figure}

Better, then, to turn to higher-resolution results.
\citet{torrejon_radial_2015}, working with \chandra-\hetgs, resolved the iron
line complex in eclipse into narrow (unresolved in the \hetgs, $\sigma \lesssim
0.005$\,\AA) \ion{Fe}{25} and neutral Fe~K$\alpha$ and K$\beta$ features. With
this in mind, we fitted the eclipse spectrum with three Gaussians with their
energies fixed to the \chandra values and their widths fixed to 0.01\,keV.
This obtains an acceptable fit ($\chi^{2}_{\rm red} = 1.12$ for the eclipse
spectrum), although compared to the two-line fit, the red wing of the iron line
is underfitted. A comparison of the fit with a narrow Fe~K$\alpha$ and a broad
6.6\,keV line to the fit with three narrow lines is displayed in
Figure~\ref{fig:compare_lines}. In this three-line fit, the iron
K$\alpha$/K$\beta$ ratio is $\sim$0.25, and is consistent within errors with
the $\sim$0.13 expected for neutral or low-ionization iron. Unfortunately, the
ingress and egress spectra are too low-exposure to provide meaningful
constraints on any changes in the lines as the pulsar moves into and out of
eclipse, typically only offering upper limits on the line fluxes and equivalent
widths. Nonetheless, our results are consistent with the \chandra values
obtained by \citet{torrejon_radial_2015}.

There is an additional dip in the eclipse residuals around $22$\,keV,
coincident with the CRSF (see panel c in Figure~\ref{fig:ecl}). Given that the
emission seen in eclipse is primarily scattered emission from the neutron star, this is not
entirely unexpected. Our fits find similar parameters in ingress, eclipse, and
egress to the out-of-eclipse CRSF. This feature remains present in eclipse even
with different choices of background region. While it is not strongly detected
--- simulations place the significance of the feature at $\sim$2.2$\sigma$ ---
this does motivate checking whether the pre-eclipse CRSF measurement is
influenced in any way by this feature. Thus, we fit the pre-eclipse spectrum
with the \texttt{mplcut} continuum using the eclipse spectrum as the
background.  This had little to no effect on the continuum and CRSF parameters.
The only notable change is that the iron lines appear significantly dimmer when
the eclipse is used as the background, due to the lines' relative brightness in
the eclipse spectrum.

% Table produced with maketable_pcf_compare.sl
\begin{deluxetable*}{lrrrrr}
  \tabletypesize{\footnotesize}
  \tablewidth{0pt}
  \tablecaption{Spectral fits for ingress, eclipse, and egress, fitted with a partial-covering absorber, a CRSF, and three iron lines\label{tab:pcf}}
  \tablehead{\colhead{}&\multicolumn{2}{c}{Ingress}&\colhead{Eclipse}&\multicolumn{2}{c}{Egress}\\
             \colhead{}&\colhead{3}&\colhead{4}&\colhead{5--12}&\colhead{13}&\colhead{14}}
  \startdata
    Flux\tablenotemark{a} ($10^{-10}$\,\eflx) & $6.48^{+0.23}_{-0.21}$ & $4.3^{+0.7}_{-0.4}$ & $0.33^{+0.08}_{-0.04}$ & $8.7\pm0.4$ & $5.74^{+0.16}_{-0.15}$ \\ 
    $C_{\rm FPMB}$ & $1.045 \pm 0.026$ & $1.01^{+0.05}_{-0.04}$ & $1.06 \pm 0.03$ & $1.017 \pm 0.021$ & $1.012 \pm 0.021$ \\
    \midrule
    $N_{\rm H}$\tablenotemark{b} ($10^{22}$\,cm$^{-2}$) & $(14.7)$\tablenotemark{c} & $(14.7)$ & $(14.7)$ & $(14.7)$ & $(14.7)$ \\ 
    $N_{\rm H,pc}$ ($10^{22}$\,cm$^{-2}$) & $76^{+9}_{-8}$ & $208^{+30}_{-22}$ & $190^{+120}_{-70}$ & $124^{+10}_{-9}$ & $47^{+9}_{-8}$ \\ 
    $f_{\rm pc}$ & $0.867^{+0.020}_{-0.018}$ & $0.950\pm0.010$ & $0.47^{+0.13}_{-0.08}$ & $0.900\pm0.009$ & $0.69^{+0.05}_{-0.04}$ \\ 
    $\Gamma$ & $(1.19)$ & $(1.19)$ & $(1.19)$ & $(1.19)$ & $(1.19)$ \\ 
    $E_{\rm cut}$ (keV) & $13.2^{+0.8}_{-0.7}$ & $25.3^{+1.5}_{-7.6}$ & $12.5^{+1.2}_{-5.8}$ & $24.5^{+0.9}_{-2.0}$ & $13.6\pm0.7$ \\ 
    $E_{\rm fold}$ (keV) & $11.8^{+0.7}_{-3.1}$ & $6.3^{+1.9}_{-1.0}$ & $13.9^{+2.2}_{-2.1}$ & $6.6^{+0.9}_{-0.5}$ & $12.9\pm0.8$ \\ 
    \midrule
    $E_{\rm fund}$ (keV) & $21.4^{+0.5}_{-0.9}$ & $23.6^{+0.9}_{-2.9}$ & $21.8^{+1.1}_{-0.8}$ & $23.6^{+0.5}_{-0.9}$ & $21.7\pm0.5$ \\ 
    $\sigma_{\rm fund}$ (keV) & $2.6^{+0.7}_{-0.5}$ & $4.9^{+1.1}_{-1.2}$ & $2.2^{+1.7}_{-0.9}$ & $4.6^{+0.5}_{-0.6}$ & $2.2^{+0.6}_{-0.5}$ \\ 
    $\tau_{\rm fund}$ (keV) & $0.53\pm0.09$ & $1.29^{+0.22}_{-0.55}$ & $0.41^{+0.15}_{-0.13}$ & $1.25^{+0.09}_{-0.13}$ & $0.58^{+0.10}_{-0.09}$ \\ 
    \midrule
    Fe K$\alpha$ energy\tablenotemark{d} (keV) & $(6.40)$ & $(6.40)$ & $(6.40)$ & $(6.40)$ & $(6.40)$ \\ 
    Fe K$\alpha$ area ($10^{-4}$\,\pflx) & $< 2.34$ & $6^{+5}_{-4}$ & $1.50^{+0.61}_{-0.29}$ & $2.5^{+2.1}_{-2.0}$ & $< 1.93$ \\ 
    Fe K$\alpha$ EW (eV) & $< 64.9$ & $260^{+140}_{-130}$ & $870\pm100$ & $50\pm40$ & $< 64.1$ \\ 
    \ion{Fe}{25} energy (keV) & $(6.68)$ & $(6.68)$ & $(6.68)$ & $(6.68)$ & $(6.68)$ \\ 
    \ion{Fe}{25} area ($10^{-4}$\,\pflx) & $< 1.98$ & $< 4.71$ & $0.62^{+0.29}_{-0.16}$ & $< 2.23$ & $< 1.45$ \\ 
    \ion{Fe}{25} EW (eV) & $< 58.1$ & $< 197$ & $380^{+90}_{-80}$ & $< 48.3$ & $< 50.7$ \\ 
    Fe K$\beta$ energy (keV) & $(7.06)$ & $(7.06)$ & $(7.06)$ & $(7.06)$ & $(7.06)$ \\ 
    Fe K$\beta$ area ($10^{-4}$\,\pflx) & $1.1\pm1.1$ & $2.6^{+1.9}_{-1.6}$ & $0.35^{+0.18}_{-0.11}$ & $< 1.51$ & $0.8\pm0.8$ \\ 
    Fe K$\beta$ EW (eV) & $< 66.5$ & $120\pm80$ & $230\pm60$ & $< 34.7$ & $< 56.4$ \\ 
    \midrule
    $\chi^{2}_{\rm red}$ & 1.08 (438) & 0.92 (288) & 1.10 (362) & 0.99 (518) & 1.01 (474) \\ 
  \enddata
  \tablenotetext{a}{Unabsorbed flux in 3--50\,keV band.}
  \tablenotetext{b}{Fixed to $N_{\rm H}$ measured in pre-eclipse spectrum.}
  \tablenotetext{c}{Values in parentheses were frozen during fitting.}
  \tablenotetext{d}{Fe K$\alpha$, K$\beta$, and \ion{Fe}{25} line widths fixed to 0.01\,keV.}
\end{deluxetable*}

\begin{figure}
  \plotone{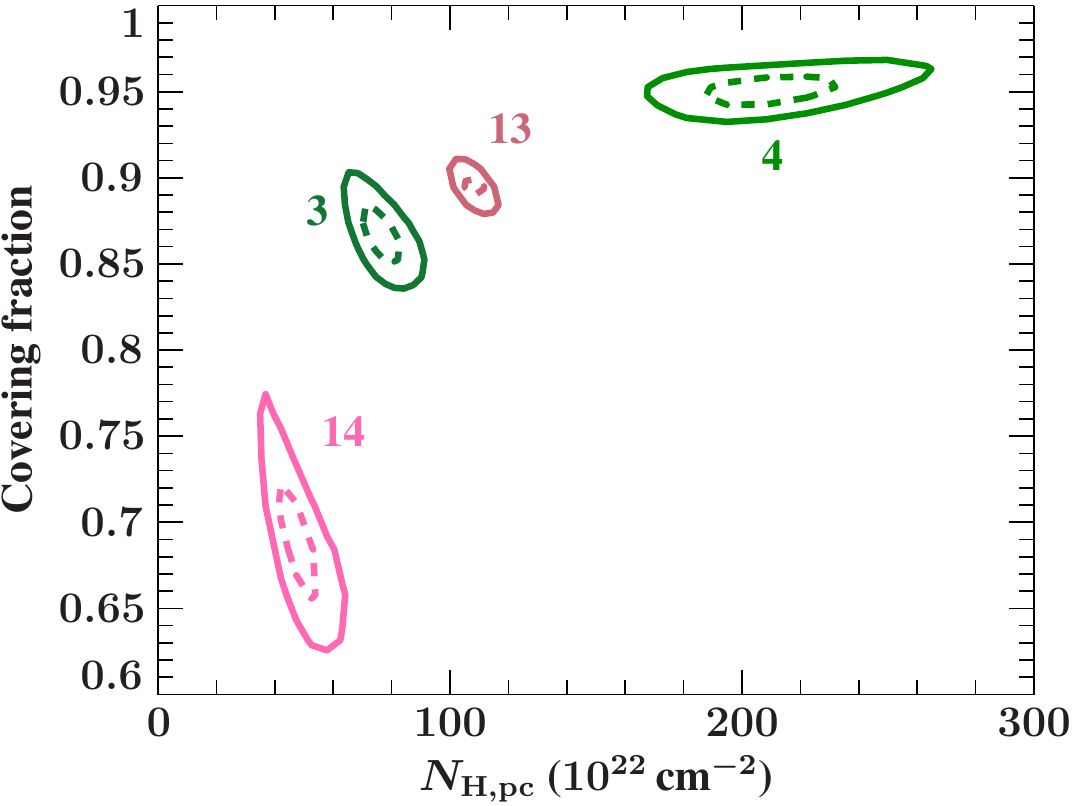}
  \caption{2-D confidence contours for \fpc \textit{vs.} \nhpc for
    the ingress (contours 3 and 4, green) and egress (contours 13 and 14, red).
    We plot the 68\% (dashed) and 99\% (solid) confidence contours.}
  \label{fig:pcf_contours}
\end{figure}

We have additionally computed two-dimensional confidence contours for \nhpc
\textit{vs.} \fpc for each phase of the observation; we display the contours
for the ingress and egress in Figure~\ref{fig:pcf_contours}. The 99\% contours
for each spectrum do not overlap, and the distribution and shapes of the
contours for ingress and
egress are, for the most part, orthogonal to the evolution of \nhpc and \fpc,
implying that the changes seen are real and not an artifact of correlations
inherent in the model used. The contours from the eclipse are complex and
bimodal, showing a correlation between \nhpc and \fpc, but are generally
consistent with a lower average covering fraction compared to ingress and
egress, consistent with its origin in scattered emission (see discussion in
Section~\ref{sec:discuss_ecl}). To characterize how accurately \nustar, with
its 3\,keV lower bound, can constrain \nhpc and \fpc, we also computed contours
for the pre-eclipse spectrum under the same assumptions used for the ingress,
eclipse, and egress, fixing \nh and $\Gamma$ and adding in a partial-covering
component. Note that this means we are adding a partial-covering component on
top of an already-well-fit spectrum --- this serves to characterize what kind
of partial-covering columns and covering fractions can ``hide'' below 3\,keV.
We find for the pre-eclipse contours that the covering fraction never exceeds
$0.2$ across the range of \nhpc measured in eclipse, ingress, and egress. The
significantly higher covering fractions measured in the ingress, egress, and
eclipse, combined with their well-constrained contours, lead us to conclude
that our partial-covering models are not unduly influenced by \nustar's lack of
low-energy coverage.

\section{Discussion} \label{sec:discussion}

\subsection{Cyclotron line energy} \label{sec:discuss_crsf}

\begin{figure*}
  \includegraphics[width=\textwidth]{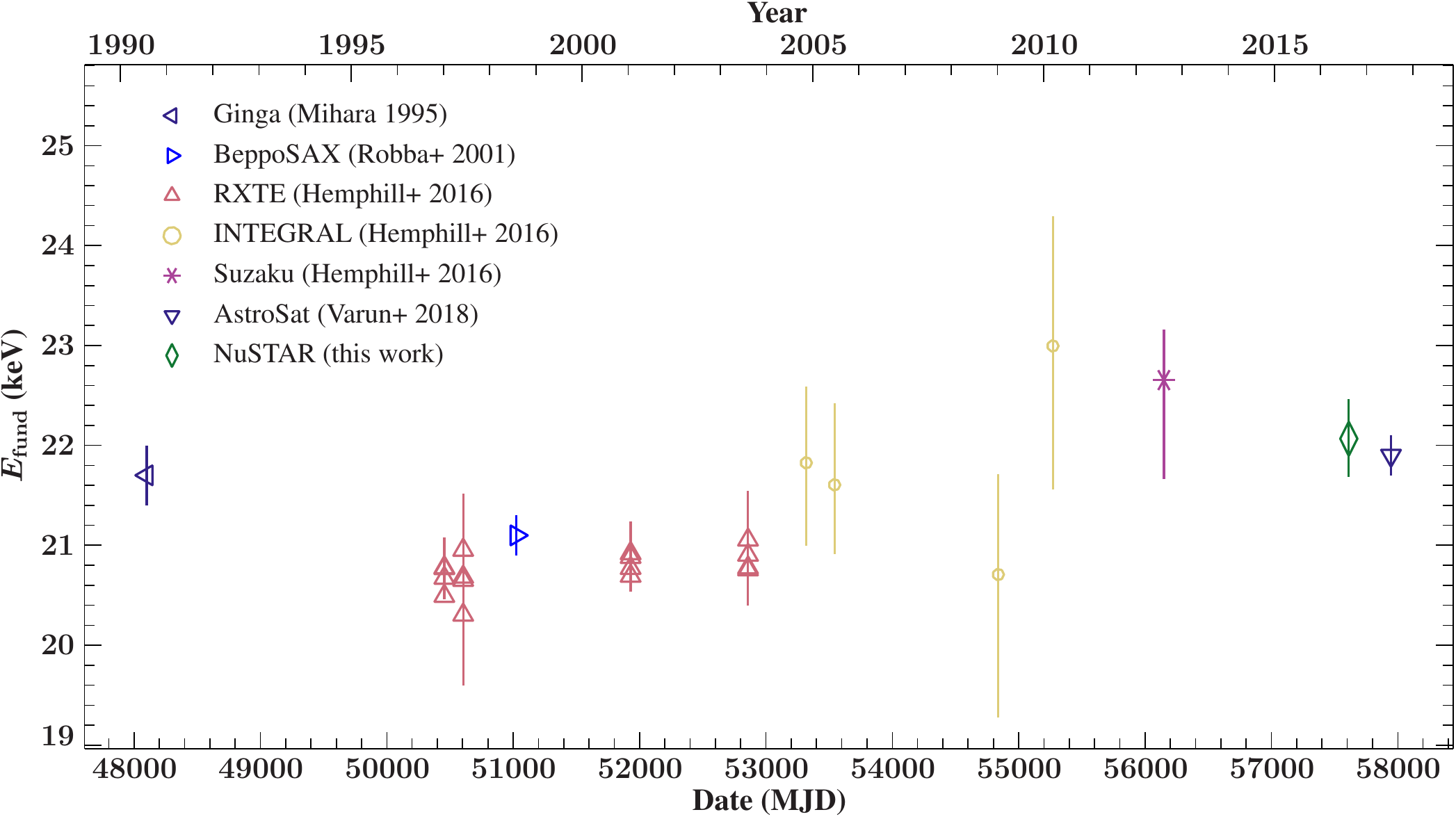}
  \caption{\fif's CRSF energy over time, as measured with \textit{Ginga} by
    \citet{mihara_observational_1995}, \textit{BeppoSAX} by
    \citet{robba_bepposax_2001}, \rxte, \suz, and \integral by
    \citet{hemphill_evidence_2016}, and \textit{AstroSat} by
    \citet{varun_probing_2018}, along with the \nustar measurement presented
    here. The \nustar measurement is from the pre-eclipse phase. All
    measurements use Gaussian optical depth profiles for the CRSF, and all but
    the \textit{Ginga} measurement use some variant of the \texttt{highecut}
    continuum.}
  \label{fig:ecyc_history}
\end{figure*}

\citet{hemphill_evidence_2016} presented evidence that the CRSF energy had
increased by $\sim$1\,keV between the early-2000s \rxte observations and the
2012 \suz observation. The \nustar measurements presented here, the first
measurement of the CRSF energy since the \suz observation, again find an elevated CRSF
energy relative to the \rxte measurements. While the \nustar measurement taken alone
is of moderate significance, the fact that we observe an increased energy in
both \suz and \nustar increases the significance to $\sim$3.6$\sigma$. While secular decays in
persistent sources have been measured in Her~X-1 \citep[and references
therein]{staubert_continued_2016} and Vela~X-1 \citep{laparola_swiftbat_2016},
\fif would be the first persistent\footnote{The transient BeXRB V~0332+53 did
experience an increase in CRSF energy between its 2015 and 2016 outbursts
\citep{vybornov_changes_2018}, but this was following the \textit{decrease}
in CRSF energy over the 2015 outburst --- V~0332+53's very high flux,
transient nature, and much shorter timescale for CRSF variations makes it
unlikely that its changes have similar origins to \fif's.} source where we
observe an \textit{increase} in CRSF energy over time. We plot the full set of
\fif CRSF measurements, including the \textit{Ginga} and \textit{BeppoSAX}
measurements of \citet{mihara_observational_1995}\footnote{see Appendix F of
\citet{mihara_observational_1995}; since this measurement uses a
Gaussian-profile CRSF and an \texttt{npex} continuum, it should only be
directly compared to the third column in Table~\ref{tab:pre} --- care should be
taken when comparing it to the rest of the work cited here due to the different
continuum model.} and \citet{robba_bepposax_2001}, as well as the recent
\textit{AstroSat} measurement by \citet{varun_probing_2018}, in
Figure~\ref{fig:ecyc_history}. The data are consistent either with a relatively
slow increase in energy from the final \rxte measurement to the most recent
\integral or \suz points, or with an abrupt jump in energy \textit{circa} 2009.
Of course, the small number of measurements, large error bars, and relatively
low significance of the measured increase in energy make drawing strong
conclusions here risky.

The origins of such a shift are unclear. Any change in the long-term average
CRSF energy would have to be due to a \emph{local} change in the field, as the
global field is unlikely to change significantly on a decadal timescale. The
decay of Her~X-1's CRSF energy is discussed in some detail in
\citet{staubert_long-term_2014}, who give a list of possible mechanisms for the
observed trend. \fif is obviously a somewhat different source from Her~X-1: it
has a lower accretion rate and no (detectable) correlation between its CRSF
energy and luminosity, although, as noted earlier and as stated in
\citet{hemphill_evidence_2016}, reasonable assumptions place \fif in the
\textit{supercritical} regime, where we expect a negative CRSF-luminosity
correlation. But the basic question is essentially the same: what is the
long-term behavior of the accretion mound under accretion, and what is the
response of the magnetic field to that behavior? 

As pointed out by \citet{staubert_long-term_2014}, there is relatively little
theoretical work suitable for this subject, due to the intractability of the
problem. CRSF production as a function of geometry is currently being
investigated \citep[this will be covered in Falkner et al., in prep., working
from the models of][]{schwarm_cyclotron_2017,schwarm_cyclotron_2017a}. But
precisely what that geometry should be, and how it will change under accretion,
is at the moment an unsolved problem. \citet{mukherjee_mhd_2013-1} have
probably the most advanced simulation results, looking at quasi-static
accretion mounds on timescales of milliseconds, which at least confirm that the
accretion mound can have a sizeable influence over the magnetic field lines.
However, 3D MHD simulations cannot be practically carried out for the
multi-year timescales we observe here.

In any case, in \fif, we have an increase in CRSF energy, unlike the decay
observed in Her~X-1. This at least suggests that, when considering the
mechanism that drives the evolution of Her~X-1's CRSF, we should look to
processes that can also produce the reverse trend under different
circumstances. An intriguing aspect of this issue is the behavior of Her~X-1's
CRSF energy between 1991 and 1994, when it appears to have jumped by
$\sim$7\,keV. \citet{staubert_long-term_2014} suggested that this indicates
that Her~X-1's CRSF energy has a natural ``floor'' of $\sim$37\,keV, and indeed
\citet{staubert_inversion_2017} report evidence that the decay has reversed and
the CRSF energy has begun to increase. This is interesting in comparison to the
behavior of \fif's CRSF, which is certainly \textit{consistent} with a
relatively short-timescale increase. However, \fif's increase is considerably
smaller than what was observed in Her~X-1\ in the early 1990s --- an increase
of $\sim$5\% compared to the $\sim$20\% seen in Her~X-1. Additionally, with the
relatively small number of measurements and large uncertainties, the data do
not specifically prefer a sudden increase over a simple linear increase. Due to
the statistically middling significance of the increase in the first place, it
is probably best to avoid too much speculation beyond this point. New
observations every few years will be needed to drive down the uncertainties
here.

As a final note, it may be tempting to hypothesize that the increase in CRSF
energy is associated with \fif's torque reversals, which took place in 1990
(close to the \textit{Ginga} measurement) and 2008 (between the \rxte and \suz
measurements). However, this would be a statistically dangerous leap to make,
as the statistical significance of the \textit{Ginga} measurement over the \rxte
CRSF energies is highly doubtful --- \citet{hemphill_evidence_2016} placed its
significance at $\sim$2.2$\sigma$, and this was without taking into account any
systematic shift in CRSF energy due to the different continuum models used (see
Table~\ref{tab:pre} --- the \texttt{npex} model, which was used by
\citeauthor{mihara_observational_1995}, may find systematically higher CRSF
energies than the \texttt{plcut} model, even when the CRSF model is the same).

\subsection{The X-ray eclipse} \label{sec:discuss_ecl}

This \nustar observation provides high-quality spectra of \fif as it passes
through eclipse, with data extending out to $\sim$35\,keV in eclipse. The
broad-band flux drops by $\sim$97\% in eclipse, similar to the 97\% drop in
flux reported by \citet{torrejon_radial_2015} in \chandra-\hetgs observations,
while the iron line fluxes remain consistent with their pre-eclipse levels.
Weak pulsations visible in the folded eclipse lightcurve are in phase with the
pulsations seen in the pre-eclipse lightcurve. This supports the notion put
forth by \citet{torrejon_radial_2015} that a significant fraction of the
scattered emission seen during eclipse comes mostly from relatively close to
the neutron star and donor (significantly less than $526$\,lt-s).

Under the assumption that changes in absorption are the sole driver of spectral
changes across the eclipse, our fits with a partial-covering absorber find an
increase in both \nhpc and \fpc during ingress, and a decrease in both
quantities during egress. This is in line with what one would expect if
\qvnor's wind is composed of small, dense clumps --- as the pulsar moves into the
limb of \qvnor, the line of sight is obscured by more and more clumps. The
change in both quantities also reflects the asymmetric eclipse profile, with
\nhpc and \fpc dropping off more gradually in egress than they rose during
ingress. The eclipse spectrum, on the other hand, is well-fit by a single
absorber, and when fit with a partial-covering model, finds a lower covering
fraction than the ingress or egress. In both cases, though, the absorber's
parameters are poorly constrained. This is due to the fundamentally different
nature of the eclipse spectrum: the ingress and egress spectra are still
dominated by the direct emission from the neutron star, and thus sample only a
single, highly-absorbed line of sight, while the eclipse spectrum is produced
by scattered emission, which samples a wider range of absorbing columns. As
noted in Section~\ref{sec:ecl}, the single-absorber fits to the eclipse find
upper limits on \nh of $\sim 5\times 10^{22}$\,cm$^{-2}$, which is consistent
with the ISM absorption towards this source.

The full coverage of the X-ray eclipse gives us a new mid-eclipse
time to compare to previous measurements. This has provided better evidence
of orbital period decay in \fif, where previous efforts had typically found
$\dot{P}/P$ consistent with zero. Our measured value is marginally significant,
at $\dot{P}/P = \left(-0.95 \pm 0.37\right) \times 10^{-6}$\,yr$^{-1}$, but
this represents a large improvement over previous results. It is also
consistent with previous measurements
\citep[e.g.,][]{corbet_orbit_1993,rubin_observation_1997,clark_orbit_2000,baykal_recent_2006,falanga_ephemeris_2015}
and is of the same sense and a similar order of magnitude as in other sources
with measured orbital period decay \citep[for a number of examples,
see][]{falanga_ephemeris_2015}. The errors on our value are in relatively part
due to the relatively large uncertainties on $\omega$ and $\dot{\omega}$; a
review of the available X-ray datasets with an eye towards improving the
orbital solution even further might drive these errors down somewhat. However,
for now, it is at least relatively simple to address some simple cases for the
cause of the orbital period decay.

Mass loss from the donor can, in principle, drive orbital period decay, but we
find that this requires somewhat specific conditions to explain \fif's
behavior. \citet{vandenheuvel_interacting_1994}'s third case for orbital period
changes, where matter is ejected into a ring around the system, can produce
$\dot{P}/P$ of order a few $\times 10^{-6}$\,yr$^{-1}$ for a range of ring
sizes, assuming \qvnor's mass-loss rate is $8.3\times 10^{-7}$\,\msol\,yr$^{-1}$
\citep{falanga_ephemeris_2015}. This is actually quite high compared to our
measured value, so either \qvnor's mass-loss rate is lower than the estimate
given by \citeauthor{falanga_ephemeris_2015}, or this is not the mechanism of
orbital period decay.

A contrasting view of mass-loss in the context of orbital period changes can be
found by working along the same lines as \citet{kelley_orbital_1983}.
\citeauthor{kelley_orbital_1983} characterized the effect of mass-loss on the
binary in terms of the amount of orbital angular momentum carried away per unit
mass lost. We find that our measured $\dot{P}_{\rm orb}/P_{\rm orb}$ would
require that the wind carry $\sim$150 times the average angular momentum
carried by an isotropic wind (to put this in the terminology used by
\citeauthor{kelley_orbital_1983}, we find $\xi \approx 150$ for a mass-loss
rate of $8.3 \times 10^{-7}$\,\msol\,yr$^{-1}$). While clumping in the wind
will certainly inflate the angular momentum carried away by the wind
\citep[see, e.g.,][]{elmellah_accretion_2018}, an increase by this large of a
factor would require some significant channeling or streaming through the
Lagrange points, which in turn would require very favorable geometry for this
to not be seen in some sense in the X-ray profile of the orbit, which is
largely featureless aside from the eclipse.

Tidal forces and stellar evolution are, we believe, more likely culprits in any
observed orbital period decay. \citet{levine_discovery_1993}, working on the
very similar system SMC~X-1, investigated the combined effects of weak tidal friction
(equilibrium tides) and the expansion of the companion, reasoning that the
expansion would work to keep the orbital and rotational motion of the companion
desynchronized and maintain the tidal dissipation of the orbit. From
\citeauthor{levine_discovery_1993},
\begin{equation}
  \frac{\dot{P}_{\rm orb}}{P_{\rm orb}} = -\frac{\omega_{\rm d} d \ln I_{\rm d} / dt}{\omega_{\rm K}\left(\mu a^2/3I_{\rm d} - 1\right)},
\end{equation}
where $\omega_{\rm d}$ and $I_{\rm d}$ are the rotational angular velocity and
moment of inertia of the donor, respectively, $\omega_{\rm K}$ is the orbital
angular velocity, $\mu$ is the reduced mass of the system, and $a$ is the
orbital separation. The moment of inertia of \qvnor is not known, but
taking \citet{levine_discovery_1993}'s estimate of $(\mu a^2 / 3I - 1) \approx
0.4$--1.4 for SMC~X-1, we find that to reach our measured $\dot{P}_{\rm
orb}/P_{\rm orb}$, we require $\mathrm{d}(\ln I)/\mathrm{d}t$ to be in the
range $3\times 10^{-7}$--$1\times 10^{-6}$\,yr$^{-1}$ for $\omega_{\rm
K}/\omega_{\rm c} \approx 0.91$ \citep[for this see, e.g.,][who also worked by
analogy to SMC~X-1]{rawls_refined_2011,falanga_ephemeris_2015}. Somewhat
unsurprisingly, this is similar to the value reported by
\citet{levine_discovery_1993} for SMC~X-1, which they state is not
unreasonable for evolved supergiant stars.

\section{Conclusions}

We have analyzed the first \nustar observation of the high-mass X-ray binary
\fif, covering the X-ray eclipse with its ingress and egress. We have carried
out both spectral and timing analyses. For the spectral analysis, our main
conclusions are as follows:
\begin{itemize}
  \item In the pre-eclipse spectrum, the CRSF is detected at $22.0 \pm
    0.4$\,keV, consistent with the 2012 \suz results of
    \citet{hemphill_evidence_2016}.
  \item The new CRSF measurement, taken alone, is 3.1$\sigma$ above the \rxte
    measurements, while \suz and \nustar measurements combined represent a
    $3.6\sigma$ increase in energy.
  \item The eclipse spectrum shows a complex iron line structure, which is
    consistent with the three narrow Fe lines detected with \chandra
    \citep{torrejon_radial_2015}.
  \item Line-of-sight absorption to the source increases dramatically during
    eclipse ingress, but is overall lower (and consistent with zero local
    absorption) during the eclipse.
\end{itemize}
The timing study is mainly concerned with updating the orbital solution and
estimating the rate of change of the orbital period:
\begin{itemize}
  \item The midpoint of the X-ray eclipse is found at MJD $57612.401 \pm
    0.013$, approximately $0.1$\,days early relative to the ephemeris of
    \citet{falanga_ephemeris_2015}.
  \item The new mid-eclipse time allows us to update the orbital period and
    epoch, with $P_{\rm orb} = 3.72831(2)$\,d and $T_{\pi/2} = 57612.53(5)$.
  \item \fif's measured rate of apsidal advance can only account for $\sim$24\%
    of the change in the eclipse-to-eclipse period.
  \item We produce a new constraint on the true orbital period derivative, with
    $\dot{P}_{\rm orb}/P_{\rm orb} = \left(-0.95 \pm 0.37\right)\times
    10^{-6}$\,yr$^{-1}$.
\end{itemize}

Both of these areas of study are concerned with the long-term evolution of the
source, and as such will be greatly aided by future observations.

\acknowledgements{
  This work was supported by NASA grant NNX17AC33G. We have made extensive use
  of the ISISscripts, a collection of ISIS routines provided by ECAP/Remeis
  observatory and MIT (\url{http://www.sternwarte.uni-erlangen.de/isis/}). The
  colors used in plots are selected based on Paul Tol's recommendations found
  in SRON Technical Note SRON/EPS/TN/09-002
  (\url{https://personal.sron.nl/~pault/data/colourschemes.pdf}) Some
  computations in Section~\ref{sec:timing} were carried out using
  \textit{Mathematica}, version 11.3. We thank Alan Levine and Nevin Weinberg
  for useful discussions, and the anonymous referee for their useful
  suggestions.
}

\facilities{NuSTAR}
\software{ISIS \citep{houck_isis:_2000},
          HEASOFT,
          Mathematica}

\bibliographystyle{aasjournal}
\bibliography{refs}

\end{document}